\documentclass[aip,reprint,a4paper]{revtex4-1}
\usepackage[a4paper,left=1.8cm,right=1.8cm,top=2.2cm,bottom=2.2cm]{geometry}
\usepackage{tikz}
\usepackage{tabularx}
\usepackage{multirow}
\usepackage{amsmath}
\usepackage{amsfonts}
\usepackage{bm}
\usepackage{hyperref}
\hypersetup{colorlinks=true,linkcolor=blue,urlcolor=blue,citecolor=blue}

\usepackage{microtype}
\usepackage{graphicx}
\usepackage{dcolumn}
\usepackage{bm}
\usepackage{mdframed}
\usepackage[utf8]{inputenc}
\usepackage[T1]{fontenc}
\usepackage{mathptmx}
\usepackage{etoolbox}
\usepackage{booktabs}
\usepackage{array}
\usepackage{tabularx}
\usepackage{ragged2e}

\draft

\begin{document}

\title{Direct inference of viscoelastic memory from chirp rheometry via physics-informed Gaussian processes}

\author{Isaac Y. Miranda-Valdez}
\email[]{isaac.miranda.valdez@alumni.aalto.fi}
\affiliation{Department of Applied Physics, School of Science, Aalto University, P.O.\ Box 15600, Espoo, FI-00076 Aalto, Finland}
\affiliation{Deltan Labs, P.O.\ Box 11000, Espoo, FI-00076 Aalto, Finland}

\author{Juha Koivisto}
\affiliation{Department of Applied Physics, School of Science, Aalto University, P.O.\ Box 15600, Espoo, FI-00076 Aalto, Finland}

\author{Mikko J. Alava}
\email[]{mikko.alava@aalto.fi}
\affiliation{Department of Applied Physics, School of Science, Aalto University, P.O.\ Box 15600, Espoo, FI-00076 Aalto, Finland}

\date{\today}

\begin{abstract}
Soft materials remember their deformation history, and identifying that memory from experiments is essential for predicting how these materials behave under real-world loading conditions. Chirp rheometry has recently emerged as a way to accelerate this characterization, compressing hours of conventional measurement into seconds and yielding thousands of stress-strain pairs per experiment. That density is then largely discarded: the standard pipeline reduces the record to a handful of frequency-domain estimates before any constitutive model is fitted. We introduce a physics-informed Gaussian process framework that infers the material's constitutive law directly from the raw time-domain record of a single chirp, selecting among candidate memory kernels and parametrizing the selected one without any intermediate signal processing step. Because the framework infers the memory kernel rather than the specific waveform used during training, it predicts the response to deformation histories it never saw, without retraining. The method also resolves material evolution within a single chirp directly in the time domain.
\end{abstract}


\maketitle

\section{Introduction}

Soft materials store memory. In a gel, a polymer solution, or a biological tissue, the stress at any instant is not determined by the current strain alone but by a weighted integral over the entire deformation history---a hereditary response whose character distinguishes, for example, a Maxwell fluid from a critical gel. Inferring this memory precisely, and knowing how much the available experimental data constrain it, is one of the central problems of modern rheology.~\citep{mangal_saadat_jamali_2025, mahmoudabadbozchelou_kamani_rogers_jamali_2024, lennon_mckinley_swan_2023}

The standard route is small amplitude oscillatory shear (SAOS); the material is probed at one frequency at a time, and the frequency-dependent storage and loss moduli, $G^{\prime}(\omega)$ and $G^{\prime\prime}(\omega)$, are assembled into a spectrum from which a constitutive model is then fitted.~\citep{ghiringhelli_roux_bleses_galliard_caton_2012, ricarte_shanbhag_2024} SAOS is mature and reproducible, but it has two practical liabilities. First, a full frequency sweep can take tens of minutes to hours, which is problematic for evolving systems---forming gels, aging suspensions, maturing biological scaffolds---that change measurably on the timescale of the measurement.~\citep{das_vadillo_perego_mckinley_2026} Second, constitutive inference occurs only after spectral estimation, so any noise or finite-window artifact introduced during that reduction propagates irreversibly into the fitted model.~\citep{singh2022simultaneousfittingnonlinearlinear}

Time-resolved mechanical spectroscopy via chirp rheometry offers a compelling alternative. A strain-controlled (or stress-controlled) signal whose instantaneous frequency sweeps continuously across a prescribed bandwidth excites many relaxation modes simultaneously, compressing the material response across multiple decades into a single experiment lasting seconds, typically the duration being set primarily by the period of the lowest frequency in the sweep.~\citep{geri_keshavarz_divoux_clasen_curtis_mckinley_2018,athanasiou_geri_roose_mckinley_petekidis_2024,perego_vadillo_mills_das_mckinleyfrs_2025, vadillo_2026_torch} This changes not only the duration of the experiment but also the character of the data: a single chirp sampled at typical rheometer rates yields $10^{3}$--$10^{4}$ stress--strain pairs spanning several decades of excited timescales, a data density unusual in a field whose established protocols remain time-intensive and comparatively sparse.~\citep{vadillo_2026_torch} The demand for such records is growing, as physics-informed machine learning and digital-twin approaches to material modeling require dense, high-resolution measurements to train and validate predictive models.~\citep{vadillo_2026_torch, lennon_mckinley_swan_2023}

Despite this efficiency, the standard chirp-analysis pipeline still mimics the SAOS workflow. Specifically, the measured time series is windowed and Fourier-transformed to estimate $G^{\prime}(\omega)$ and $G^{\prime\prime}(\omega)$, and a constitutive model is fitted only after that reduction to compactly describe the material. For short or noisy records---precisely the cases where chirp rheometry is most attractive---the Fourier estimates can be sparse, scattered, and strongly influenced by finite-window artifacts.~\citep{HUDSONKERSHAW2024105307, waeterloos_mckinley_clasen_2025} This reduction is severe: a record of $10^{3}$--$10^{4}$ samples is compressed into only a few tens of $(G',G'')$ estimates, so most of the measured information reaches the constitutive fit only through that summary. The result is not merely data compression but an inferential bottleneck, because any bias or instability introduced during spectral estimation propagates directly into the recovered constitutive law.

Exploiting that density calls for methods that can learn from a dense time-domain record, and machine learning is increasingly applied to rheology.~\citep{qi_yin_pang_lin_wang_zhu_chen_sha_bai_2026, lennon_mckinley_swan_2023b, naoki_2020, saadat_mangal_jamali_2023} Unconstrained data-driven models, however, face a fundamental obstacle; namely, a regressor trained on one waveform may not generalize to a different deformation protocol.~\citep{lennon_mckinley_swan_2023, mangal_saadat_jamali_2025} Without the hereditary structure of a constitutive law, such models are waveform predictors rather than material descriptions. A notable example is rheology-informed neural networks (RhINNs), which embed a candidate constitutive model as residual constraints in the loss function and have been used to recover viscoelastic and thixotropic parameters, including model selection across a manually specified library.~\citep{saadat_mangal_jamali_2023, saadat_mahmoudabadbozchelou_jamali_2022} However, they lack predictive power: because the network's own output lacks reliable extrapolation behavior, forward predictions are generated by re-inserting the recovered parameters into a separate ordinary differential equation solver rather than by the network itself.~\citep{saadat_mangal_jamali_2023}

Here, we resolve these limitations and exploit the dense time-resolved records produced by chirp rheometry via direct probabilistic constitutive inference in the time domain. We embed the Boltzmann superposition principle into the input space of a sparse Gaussian process (GP)~\citep{hensman2013gaussian, rasmussen_2005, gardner2018gpytorch} by constructing causal convolutional features. These features are stress-like signals obtained by filtering the measured strain-rate history through candidate relaxation kernels. The GP maps these features to the measured stress, enabling direct inference of material parameters, uncertainty, and model evidence from the raw time-domain record. Because the learned memory kernel is a constitutive property of the material, independent of the applied waveform, the GP, once trained, can be evaluated on unobserved deformation histories, predicting the stress response directly.

We demonstrate five capabilities of the framework: \textbf{(i)}~from a chirp of a few seconds on a wormlike micellar solution, evidence-based model selection identifies a Fractional Maxwell liquid at its Maxwell limit and recovers a relaxation time whose crossover frequency lies below the excited band; \textbf{(ii)}~from a short, noisy chirp on a cellulose nanofiber hydrogel, the method recovers the same scale-free critical-gel memory as a conventional frequency sweep sixty times longer; \textbf{(iii)}~from a chirp on a photopolymer resin, the GP apportions the stress between the two parallel branches of a Fractional Kelvin--Voigt model, resolving a permanent elastic network alongside a power-law dissipative fraction with quantified uncertainty in the decomposition; \textbf{(iv)}~the same chirp-trained GP predicts the stress response to an independent chirp of different bandwidth, duration, and initial phase, resolving a terminal crossover that lies outside the training band; \textbf{(v)}~from a single chirp on a UV-curing acrylate, the time-resolved GP sensitivity detects the onset of material mutation as a violation of time-translation invariance. Conceptually, the framework replaces the conventional sequence of spectral estimation followed by constitutive fitting with a single physics-informed probabilistic inference problem posed directly on the raw chirp response.

\section*{Results}

\subsection*{Inferring viscoelastic memory directly from chirps}

In the linear viscoelastic regime, the Boltzmann superposition principle gives
\begin{equation}
  \sigma(t) = \int_{0}^t G(t-t')\,\dot\gamma(t')\,\mathrm{d}t'
  \label{eq:boltzmann}
\end{equation}
where $\sigma(t)$ is the measured shear stress, $G(t)$ is the relaxation modulus, and $\dot\gamma(t')=\mathrm{d}\gamma/\mathrm{d}t'$ is the strain rate; the causal integral runs over all past times $t'<t$. We represent a candidate memory as a weighted sum of kernel shapes $\phi$,
\begin{equation}
  G(t;\bm\Theta) = \sum_{i=1}^n c_i\,\phi_i(t;\bm\kappa)
  \label{eq:kernel-decomp}
\end{equation}
where $\bm\kappa$ contains the memory-shape parameters (\textit{e.g.,} relaxation times $\tau_c$, fractional exponents $\alpha$, $\beta$), $c_i$ are scalar prefactors (\textit{e.g.,} shear moduli, viscosities, and quasi-properties) that linearly connect stress to strain\citep{bonfanti_kaplan_charras_kabla_2020}, and $\bm\Theta=(\bm\kappa,\mathbf{c})$ is the full parameter set. Substituting Eq.~\eqref{eq:kernel-decomp} into Eq.~\eqref{eq:boltzmann} yields
\begin{equation}
  \sigma(t) = \sum_{i=1}^n c_i\,x_i(t;\bm\kappa) \quad x_i(t;\bm\kappa) = \int_{0}^t \phi_i(t-t';\bm\kappa)\,\dot\gamma(t')\,\mathrm{d}t'
  \label{eq:features}
\end{equation}

Each $x_i(t)$ is a causal convolutional feature: the stress that would result if the material's memory were governed solely by kernel component $\phi_i$. The feature vector $\mathbf{x}(t)=[x_1(t),\ldots,x_n(t)]^\top\in\mathbb{R}^n$ replaces the raw time coordinate as the GP input, encoding the full deformation history through a physically motivated filter.

A GP is assigned over the stress--feature map,
\begin{align}
  \sigma(t) &= f\bigl(\mathbf{x}(t)\bigr)+\varepsilon(t), \quad \varepsilon(t)\sim\mathcal{N}(0,\mathrm{var}^2) \notag\\
  f(\mathbf{x}) &\sim \mathcal{GP}\!\bigl[m(\mathbf{x}) ,K(\mathbf{x},\mathbf{x}')\bigr]
  \label{eq:gp}
\end{align}
Here $\varepsilon(t)$ is the observation noise with variance $\mathrm{var}^2$; $m(\mathbf{x})$ is the prior mean function (set to zero after feature standardization); and $K(\mathbf{x},\mathbf{x}')$ is the GP covariance kernel, which controls the smoothness of the learned stress--feature map. The posterior predictive mean of $\sigma(t)$ given the training data is denoted $\bar{\sigma}(t)$. We train all model parameters jointly using Adam by maximizing the evidence lower bound (ELBO)~\citep{pmlr-v5-titsias09a, hensman2013gaussian}, applying sigmoid and softplus reparameterizations to the shape parameters $\bm{\kappa}$ (see Supplementary Information, Sec.~S3.3).

\begin{figure*}[t]
    \centering
    \includegraphics[width=\linewidth]{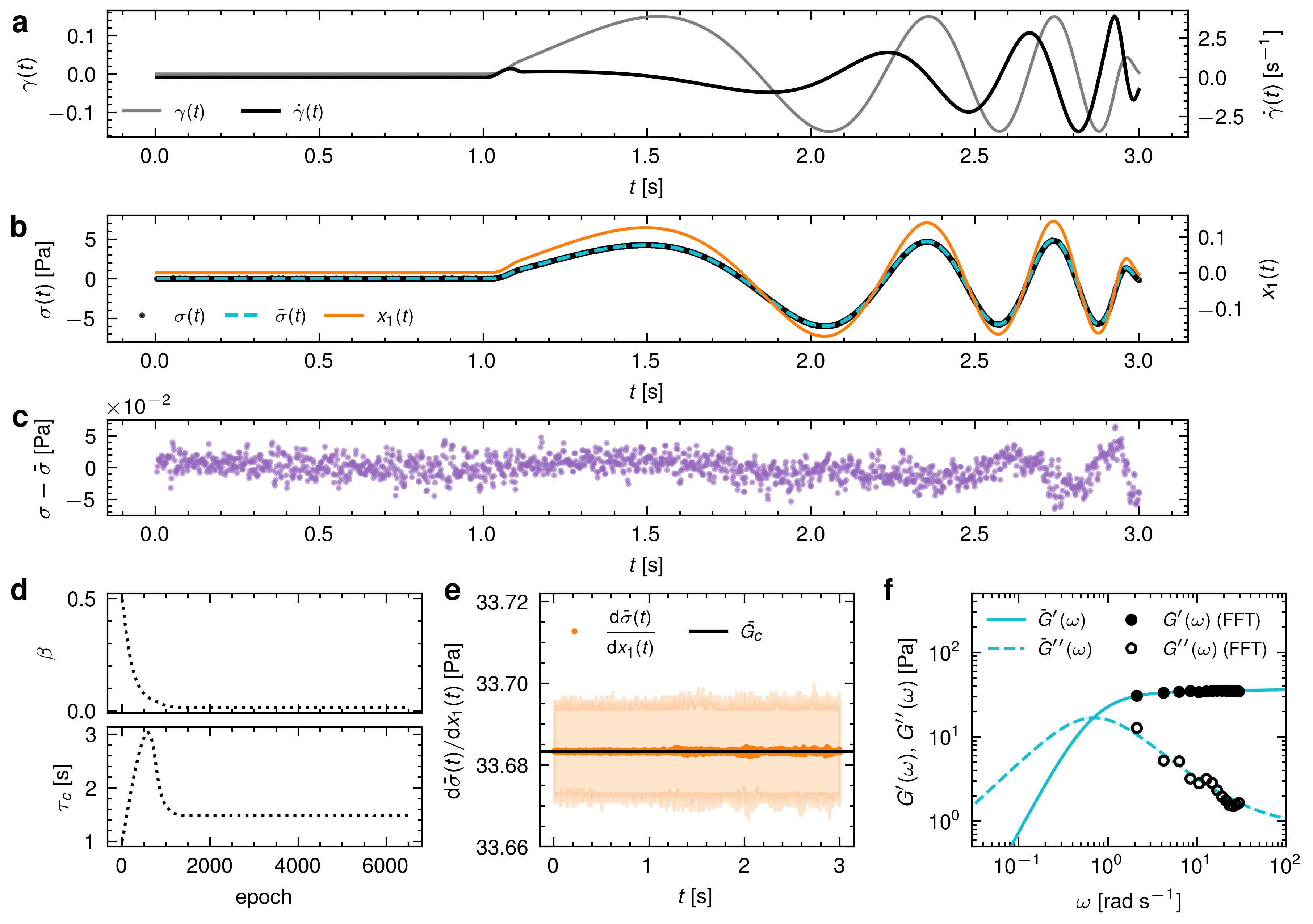}
    \caption{Time-domain, physics-informed Gaussian-process analysis of a chirp rheology experiment on a wormlike micellar solution. The experimental dataset is from \citet{das_vadillo_perego_mckinley_2026}. \textbf{a} Chirp input in the time domain, showing the imposed strain $\gamma(t)$ on the left axis and strain rate $\dot{\gamma}(t)$ on the right axis. The instantaneous angular frequency spans the range $\omega \in [3,30]~\mathrm{rad\,s^{-1}}$. \textbf{b} GP fit to the measured stress $\sigma(t)$. Black markers denote the experimental data, the dashed curve is the posterior predictive mean $\bar{\sigma}(t)$, and the shaded region is the 95\% credible band, which is imperceptible because it is thinner than the line thickness. The engineered memory feature $x_1(t)$ of the Fractional Maxwell liquid (FML) model is shown on the secondary axis. \textbf{c} Residuals $\sigma(t)-\bar{\sigma}(t)$ on the same time axis, showing that the error is uniform across the sweep. \textbf{d} Training trajectories of the learned shape parameters $\beta$ and $\tau_c$, demonstrating convergence to well-defined values. \textbf{e} Time-resolved sensitivity of the predictive mean to the memory feature, $\mathrm{d}\bar{\sigma}(t)/\mathrm{d}x_1(t)$, with pointwise credible bands. The horizontal line denotes the magnitude of the posterior mean modulus $\bar{G}_c$. \textbf{f} Storage and loss moduli inferred from the learned time-domain model. Solid and dashed curves show $\bar{G}'(\omega)$ and $\bar{G}''(\omega)$, as predicted using the GP-inferred parameters in the analytical solution of the FML model in frequency, respectively, with uncertainty propagated from the GP posterior. The credible bands are thinner than the line width. Discrete markers show conventional Discrete Fourier Transform (DFT) estimates from the same chirp within the excitation window.}
    \label{fig:micelles}
\end{figure*}

To infer viscoelastic memory from a chirp dataset, we use two GP covariance kernels. The \textit{linear kernel},
$K_\mathrm{lin}(\mathbf{x},\mathbf{x}')=s_f^2\,v_0\,\mathbf{x}^\top\mathbf{x}'$,
makes the GP equivalent to Bayesian linear regression in the physics-informed feature space.~\citep{rasmussen_2005} In this case, the posterior predictive mean is
$\bar\sigma(t)=\sum_i\bar c_i\,x_i(t)$,
and the feature sensitivities $\partial\bar\sigma/\partial x_i=\bar c_i$ are constant and directly equal to the inferred constitutive prefactors.

The \textit{RBF kernel},
$K_\mathrm{RBF}(\mathbf{x},\mathbf{x}')=s_f^2\exp(-\|\mathbf{x}-\mathbf{x}'\|^2/2\ell^2)$,
allows a nonlinear stress--feature map, where $\ell$ is an isotropic length scale in memory-feature space. In that setting, the time-resolved sensitivity
$S_i(t)=\partial\bar\sigma(t)/\partial x_i(t)$
serves as a diagnostic of linear viscoelastic consistency: near constancy supports the selected kernel, whereas systematic variation indicates model mismatch, nonlinear response, or material evolution.

For measurements in the linear viscoelastic regime, we train with the linear kernel throughout, imposing linearity \emph{a priori} while optimizing the shape parameters $\bm\kappa$ jointly with the prefactor posterior. The RBF kernel is reserved for cases where linearity should not be presupposed, as in mutating materials. Full details are given in Supplementary Information, Sec.~S3.

When the constitutive class is unknown, candidate memories are compared using the Akaike~\citep{akaike1973second} and Bayesian information criteria~\citep{leonard2001bayesian},
\begin{equation}
\mathrm{AIC} = 2\mathcal{U} + 2k \qquad
\mathrm{BIC} = 2\mathcal{U} + k\ln N
\label{eq:bic}
\end{equation}
where $\mathcal{U}=-\,\mathrm{ELBO}$ is the total negative evidence lower bound over $N$ time samples, used here as a surrogate for the negative log marginal likelihood.~\citep{pmlr-v5-titsias09a, hensman2013gaussian} The parameter count $k$ includes the memory-shape parameters $\bm\kappa$, the GP posterior prefactors $c_i$, and four GP hyperparameters. In the linear-kernel model these are the constant mean $m_0$, output scale $s_f^2$, linear-kernel variance, and noise variance; in the RBF-kernel model they are $m_0$, $s_f^2$, length scale $\ell$, and noise variance. Because only differences between models fitted to the same record are meaningful, we report $\Delta\mathrm{AIC}$ and $\Delta\mathrm{BIC}$ relative to the best-scoring model. The model library and full methodology are described in Supplementary Information, Sec.~S3 and S4, as well as in Table~S1.

\subsection*{Inferring a single relaxation time in a Maxwellian material}

\begin{table*}[t]
  \centering
  \caption{Model comparison for the wormlike-micelle chirp ($N=1499$). The fit term $2\mathcal{U}$ and the complexity penalties $2k$ and $k\ln N$ are listed separately, so that the origin of each ranking is visible; $k$ counts memory-shape parameters, prefactors, and GP hyperparameters. Only differences between candidates fitted to the same record are meaningful; absolute values are given for reference. Lower $\Delta$ is better; $\Delta=0$ marks the selected model. Dashes denote parameters absent from the model. The Scott~Blair model has no characteristic modulus $\bar{G}_c$; instead, it has a quasi-property, $\mathbb{V}=29.19~\mathrm{Pa\,s}^{\alpha}$.}
  \label{tab:micelles_ic}
  \small
  \setlength{\tabcolsep}{5pt}
  \renewcommand{\arraystretch}{1.2}
  \begin{tabular}{lcccccccccccc}
    \hline
    Model & $2\mathcal{U}$ & $2k$ & $k\ln N$ & AIC & BIC & $\Delta$AIC & $\Delta$BIC & RMSE [Pa] & $\bar{G}_c$ [Pa] & $\alpha$ & $\beta$ & $\tau_c$ [s] \\
    \hline
    FML     &  $-9017.4$ & 14 & 51.2 & $-9003.4$ & $-8966.2$ & 0      & 0      & 0.0180 & 33.68 & --    & 0.014 & 1.487 \\
    FMM     &  $-9018.2$ & 16 & 58.5 & $-9002.2$ & $-8959.7$ & 1.2    & 6.5    & 0.0180 & 33.69 & 1.000 & 0.014 & 1.487 \\
    FMG     &  $-8058.0$ & 14 & 51.2 & $-8044.0$ & $-8006.8$ & 959.4  & 959.4  & 0.0366 & 35.47 & 0.931 & --    & 1.374 \\
    Maxwell &  $-7349.9$ & 12 & 43.9 & $-7337.9$ & $-7306.1$ & 1665.4 & 1660.1 & 0.0525 & 34.82 & --    & --    & 1.267 \\
    SB      &  $-2453.7$ & 12 & 43.9 & $-2441.7$ & $-2409.9$ & 6561.6 & 6556.3 & 0.3003 & -- & 0.077 & --    & --    \\
    \hline
  \end{tabular}
\end{table*}

We first apply the framework to a wormlike micellar solution, a canonical breakable soft-matter system whose linear viscoelastic response is often close to Maxwellian.~\citep{geri_keshavarz_divoux_clasen_curtis_mckinley_2018, das_vadillo_perego_mckinley_2026} The question is whether a chirp experiment lasting only two seconds contains sufficient information to recover this single-timescale memory directly from the time-domain stress response, and whether the data justify more complex generalizations.

We analyze the chirp-rheometry dataset reported by \citet{das_vadillo_perego_mckinley_2026}. The imposed strain is an optimally windowed chirp (OWCh) in the linear viscoelastic regime (see Supplementary Information, Sec.~S1) that proceeds after a 1~s quiescent interval of zero strain, with instantaneous angular frequency swept logarithmically from $\omega_1=3~\mathrm{rad\,s^{-1}}$ to $\omega_2=30~\mathrm{rad\,s^{-1}}$ over a duration $T_{\mathrm{owc}}=2\pi/\omega_1 \approx 2~\mathrm{s}$ and sampled at $500~\mathrm{Hz}$. The resulting time series of $\gamma(t)$, $\dot{\gamma}(t)$, and $\sigma(t)$ spans a decade of excited frequencies, as shown in Fig.~\ref{fig:micelles}a,b.

Rather than committing to a constitutive model in advance, we let the data arbitrate among a library of Maxwell-type memory kernels $G(t)=G_c\,\phi(t;\boldsymbol{\theta})$: the classical Maxwell model, $\phi(t;\tau_c)=\exp(-t/\tau_c)$, and its fractional generalizations---the full Fractional Maxwell model (FMM), the Fractional Maxwell gel (FMG), and the Fractional Maxwell liquid (FML)~\citep{jaishankar_mckinley_2013, song_holten-andersen_mckinley_2023}---which replace one or both of the spring and dashpot elements by springpots and thereby admit power-law broadening of the relaxation spectrum. The Scott~Blair (SB) model, a single springpot with no characteristic time, serves as a physically implausible reference: a scale-free memory is inconsistent with the single relaxation process expected in a wormlike micellar solution. Each candidate defines a memory feature through Eq.~\eqref{eq:features}; the kernels and their parameters are summarized in Table~S1 (see Supplementary Information). All candidates are trained on the same record with identical inducing-point count, early-stopping settings, and learning rates, and scored with the complete-parameter criteria of Eq.~\eqref{eq:bic}. Table~\ref{tab:micelles_ic} presents the results of the Akaike and Bayesian information criteria analysis.

Within the specified library, both criteria select the FML, with the FMM second ($\Delta\mathrm{BIC}\approx 6.5$) and the SB model decisively rejected ($\Delta\mathrm{BIC}\approx 6.6 \times10^{3}$). The selected FML kernel replaces the Hookean spring of the Maxwell model with a springpot of order $\beta$, giving the Mittag-Leffler memory
\begin{equation}
    \phi_{1}(t;\beta,\tau_c) =
    \left(\frac{t}{\tau_c}\right)^{-\beta}
    E_{1-\beta,\,1-\beta}\!\left[-\left(\frac{t}{\tau_c}\right)^{1-\beta}\right]
    \label{eq:fml_kernel}
\end{equation}
where $E_{a,b}(\cdot)$ is the two-parameter Mittag-Leffler function, which reduces to the exponential function as $\beta \to 0$. The selection is consistent with the Maxwellian expectation, and the fitted value $\beta = 0.014$ places the model near its Maxwell limit.

Small as it is, this departure from the limit is well supported. Constraining $\beta=0$ exactly, which is the classical Maxwell model, costs $\Delta\mathrm{BIC}\approx1.66\times10^{3}$ and nearly triples the root-mean-square error (RMSE), from $0.018$ to $0.053~\mathrm{Pa}$. The same value is recovered independently by a model that does not fix $\alpha$: the full FMM, given three free shape parameters, converges to $\alpha=1.000$ (i.e., the springpot $\alpha$ reduces to a dashpot element) and returns $\beta=0.014$ and $\tau_c=1.487~\mathrm{s}$, so that two differently parametrized members of the library locate the same point in parameter space and the $\Delta\mathrm{BIC}$ separating them is entirely the cost of the redundant parameter. Physically, the micellar response is Maxwellian to leading order with a weak broadening of the relaxation spectrum at short times, most apparent in the loss modulus, where allowing $\beta>0$ captures the slight curvature of $G''(\omega)$ at the upper end of the excited band. 

Fig.~\ref{fig:micelles} presents the selected FML fit. The posterior mean follows the measured stress throughout the chirp with $\mathrm{RMSE}=0.018~\mathrm{Pa}$ (Fig.~\ref{fig:micelles}b). The residuals in Fig.~\ref{fig:micelles}c deserve closer inspection because the criteria in Eq.~\eqref{eq:bic} treat the $N$ time samples as independent observations, and this record allows that assumption to be tested directly. The record opens with a $1~\mathrm{s}$ interval at zero imposed strain. There, the memory feature vanishes identically, the model predicts zero, and the residual is instrument noise alone. The residuals scatter symmetrically about zero at $\pm5\times10^{-2}~\mathrm{Pa}$, with an integrated autocorrelation time of $\tau_{\mathrm{int}}=2.7$ samples (i.e., $5.4\times10^{-3}~\mathrm{s}$). The noise therefore decorrelates within a few milliseconds, so the sampled measurements are approximately independent.

Once the chirp begins, the residuals stay within the same range of $\pm5\times10^{-2}~\mathrm{Pa}$, spanning $0.74\%$ of the local stress amplitude over the first two seconds and $0.72\%$ over the last half second, so the fit is equally good at both ends of the excited band. Their autocorrelation time, however, rises to $\tau_{\mathrm{int}}=22$ samples, or $4.4\times10^{-2}~\mathrm{s}$, an order of magnitude above the instrument value.

The residual autocorrelation time depends on the fitted kernel: the same record gives $\tau_{\mathrm{int}}=51$, $56$, and $381$ samples for the FMG, Maxwell, and SB models. Since the instrument contributes $2.7$ samples in every case, the excess correlation is attributable to constitutive misfit. The resulting ordering matches that of the information criteria, except for the FMM, whose value $\tau_{\mathrm{int}}=20.1$ is marginally below that of the selected FML. This exception is expected: the FMM converges to $\alpha=1.000$ and returns the same $\beta$, $\tau_c$, and RMSE as the FML, so the two produce nearly identical residual sequences and differ only in parameter count. Thus, the independence assumption underlying $N$ in Eq.~\eqref{eq:bic} is satisfied at the level of the measurement, while residual correlation reflects model misfit rather than sampling.

As a robustness check, we recomputed both criteria using effective sample sizes $N_{\mathrm{eff}}=N/\tau_{\mathrm{int}}$ between $27$ and $71$. The rejection of the FMG, Maxwell and SB models is unaffected. The FML and FMM, however, are no longer separated ($\Delta\mathrm{BIC}_{\mathrm{eff}}=8.6$ and $0$), which reflects a limitation of the check rather than an ambiguity in the data: the two fits occupy the same point in parameter space, so their ranking rests entirely on the penalty for the redundant parameter, and shrinking $N$ from $1499$ to a few tens shrinks the penalty. Because a criterion evaluated at a reduced sample size is less able to penalize unused parameters, we retain $N$ as the number of measured samples, which the instrument autocorrelation time justifies, and interpret the resulting differences as a parsimony ordering.

\begin{figure*}
    \centering
    \includegraphics[width=1\linewidth]{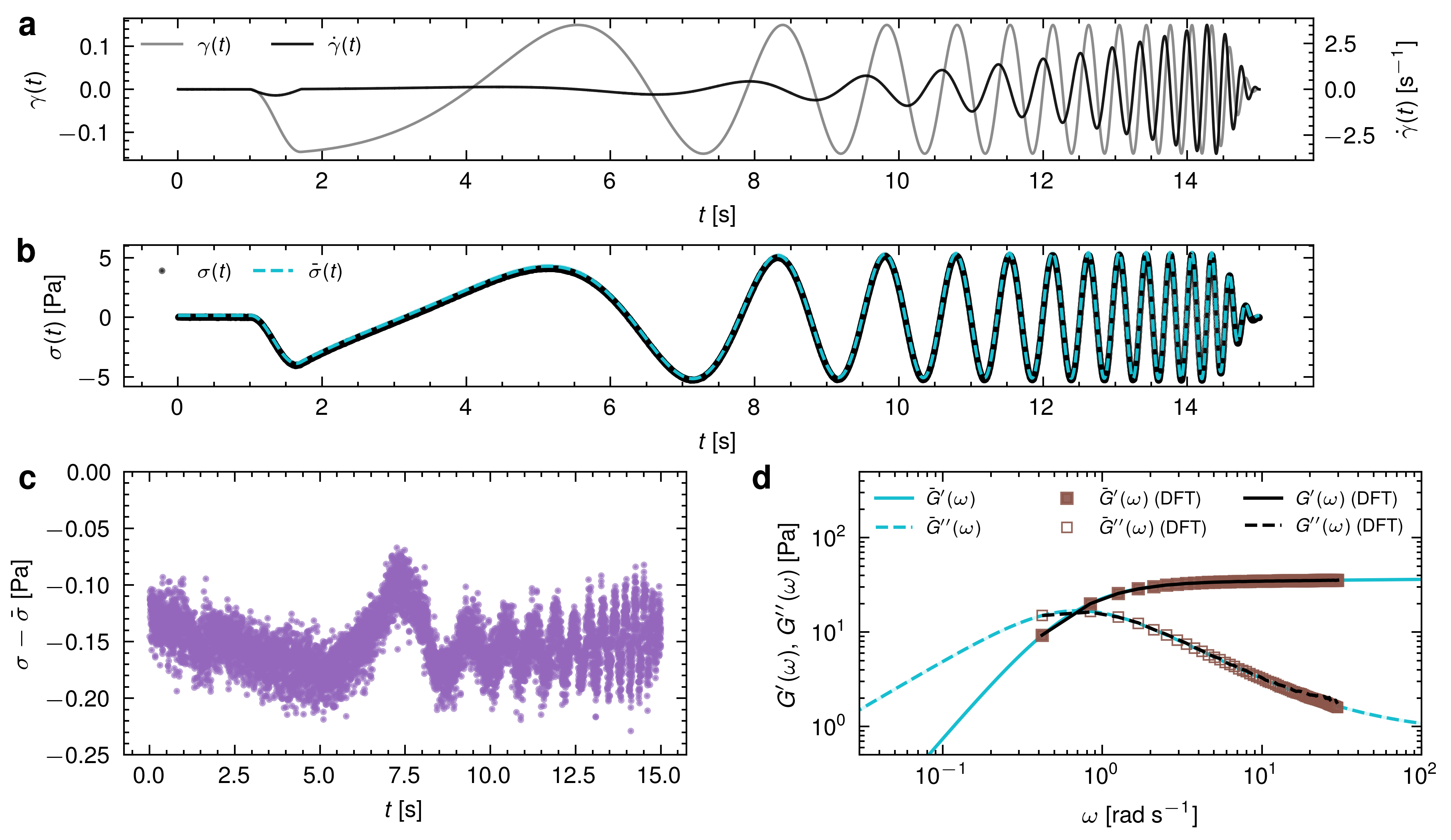}
    \caption{Gaussian process prediction and frequency‑domain validation using an OWCh strain‑controlled input different from the one presented during training. A physics‑informed GP trained on one chirp of a given material is evaluated on an independent chirp that starts with a $-90^{\circ}$ phase shift and spans the instantaneous angular‑frequency band $\omega\in[0.3,30]~\mathrm{rad\,s^{-1}}$. Panels: \textbf{a} imposed strain $\gamma(t)$ (left axis) and strain rate $\dot{\gamma}(t)$ (right axis) for the new OWCh; \textbf{b} measured stress $\sigma(t)$ (markers) and GP posterior mean $\bar{\sigma}(t)$ with 95\% band (cyan); \textbf{c} residuals $\sigma-\bar{\sigma}$; \textbf{d} frequency domain: analytical $G'(\omega)$ and $G''(\omega)$ from the GP‑inferred model (trained on the 2-s chirp, cyan lines) compared against DFT estimates from the measured OWCh (black lines) and from the GP‑predicted stress trace (brown squares). The analytic curves are also shown one decade below the excited band (down to $\omega\approx0.03~\mathrm{rad~s^{-1}}$), illustrating extrapolation beyond the OWCh window.}
    \label{fig:chirp_90}
\end{figure*}

The shape parameters themselves converge within the first few thousand epochs and remain flat thereafter (Fig.~\ref{fig:micelles}d), indicating a well-identified optimum, and the sensitivity $\mathrm{d}\bar{\sigma}(t)/\mathrm{d}x_1(t) = \bar{G}_c$ provides the magnitude of the characteristic modulus (Fig.~\ref{fig:micelles}e). In this example, linearity is imposed \emph{a priori}: the GP is equipped with the linear covariance kernel throughout training, so the stress--feature map is linear by construction and the posterior sensitivity is a single constant---the Bayesian estimate of the constitutive prefactor---while the shape parameters $\beta$ and $\tau_c$ are optimized jointly through the same variational bound. This is the appropriate choice here, since the OWCh protocol operates in the linear viscoelastic regime by design. The posterior yields
\begin{equation}
    \bar{G}_c = 33.68~\mathrm{Pa} \qquad
    95\%~\mathrm{CI} = [33.67,\,33.70]~\mathrm{Pa}
\end{equation}
We will show later that \emph{relaxing} this assumption is itself an analytical tool: with an RBF covariance the stress--feature map is free to deviate from linearity, and a time-resolved sensitivity that drifts over the record becomes a diagnostic of material mutation---a violation of time-translation invariance that a linear prefactor cannot express.

The characteristic frequency implied by the fit, $\omega_c = \tau_c^{-1} \approx 0.67~\mathrm{rad\,s^{-1}}$, lies \emph{below} the excited band: over the window $3 \leq \omega \leq 30~\mathrm{rad\,s^{-1}}$ the chirp probes only the elastic side of the relaxation process ($\omega\tau_c \approx 4.5$--$44.6$), and no spectral estimate exists at the crossover itself. A measurement lasting only a few seconds, and therefore comparable to the relaxation time it seeks to infer ($T_{\mathrm{owc}} \approx 2~\mathrm{s}$ versus $\tau_c = 1.487~\mathrm{s}$), nevertheless constrains $\tau_c$ because the time-domain stress retains memory decay that a truncated frequency-domain representation discards. Evaluating the analytical FML response in the frequency domain with the inferred parameters $(\bar{G}_c,\beta,\tau_c)$ (explicit forms in Supplementary Information, Sec.~S4) reproduces the discrete Fourier transform (DFT) estimates within the excitation band (Fig.~\ref{fig:micelles}f), confirming that the GP infers an interpretable constitutive law rather than merely interpolating the stress trace. Whether that law also holds outside the conditions it was inferred from is tested next.

\subsection*{Predicting the response to an unseen deformation history}

The defining test of constitutive inference is protocol independence: a model that truly captures a material’s rheology should predict the response to deformation histories not used for inference, provided they remain within the same constitutive regime. We test that capability directly by evaluating the chirp-trained GP of Fig.~\ref{fig:micelles} on an independent OWCh spanning $\omega \in [0.3,30]~\mathrm{rad\,s^{-1}}$, which includes the characteristic frequency $\omega_c = 0.67~\mathrm{rad\,s^{-1}}$ inferred from the 2-s training chirp. As shown in Fig.~\ref{fig:chirp_90}a, this new waveform lasts a duration of $T_{\mathrm{owc}}=14~\mathrm{s}$. To make the test more stringent, the chirp begins with a phase shift of $-90^\circ$, a feature absent from the training signal.

The residuals $\sigma-\bar{\sigma}$ of this prediction (Fig.~\ref{fig:chirp_90}c) are dominated by a constant offset of $-0.15~\mathrm{Pa}$. The same offset is already present during the 1-s quiescent interval that precedes the sweep, where the imposed strain and hence the predicted stress are identically zero and the measured stress averages $-0.13~\mathrm{Pa}$; it is therefore a transducer-baseline difference between the two records rather than an error in the constitutive prediction. After removing it, the trained GP reproduces the stress response of this new waveform with an RMSE of $0.024~\mathrm{Pa}$ against a stress amplitude of $\pm5~\mathrm{Pa}$ (Fig.~\ref{fig:chirp_90}b), comparable to the fit to the training record itself ($0.018~\mathrm{Pa}$).

Fig.~\ref{fig:chirp_90}d shows the frequency domain reached by three routes. The cyan curves are the analytical FML moduli evaluated with the parameters inferred from the two-second record; the black lines are the DFT of the measured stress of the $14~\mathrm{s}$ chirp; and the orange markers are the DFT of the stress predicted for that chirp by the trained GP. All three coincide across the excited band, including the terminal crossover near $\omega_c \approx 0.67~\mathrm{rad\,s^{-1}}$, which the extended chirp resolves directly but which lies an order of magnitude below the band the model was trained on. Fitting the FML model to the DFT of the $14~\mathrm{s}$ chirp gives $\tau_c = 1.53~\mathrm{s}$; the time-domain inference from the two-second record returns $\tau_c = 1.487~\mathrm{s}$, within $3\%$ of that reference. 

The GP has therefore learned a constitutive law rather than a waveform-specific mapping: a memory kernel inferred from only two seconds of data predicts, without refitting, both the stress response to a different deformation history and a relaxation time that the training chirp itself did not directly resolve.

\begin{figure*}[t]
    \centering
    \includegraphics[width=\linewidth]{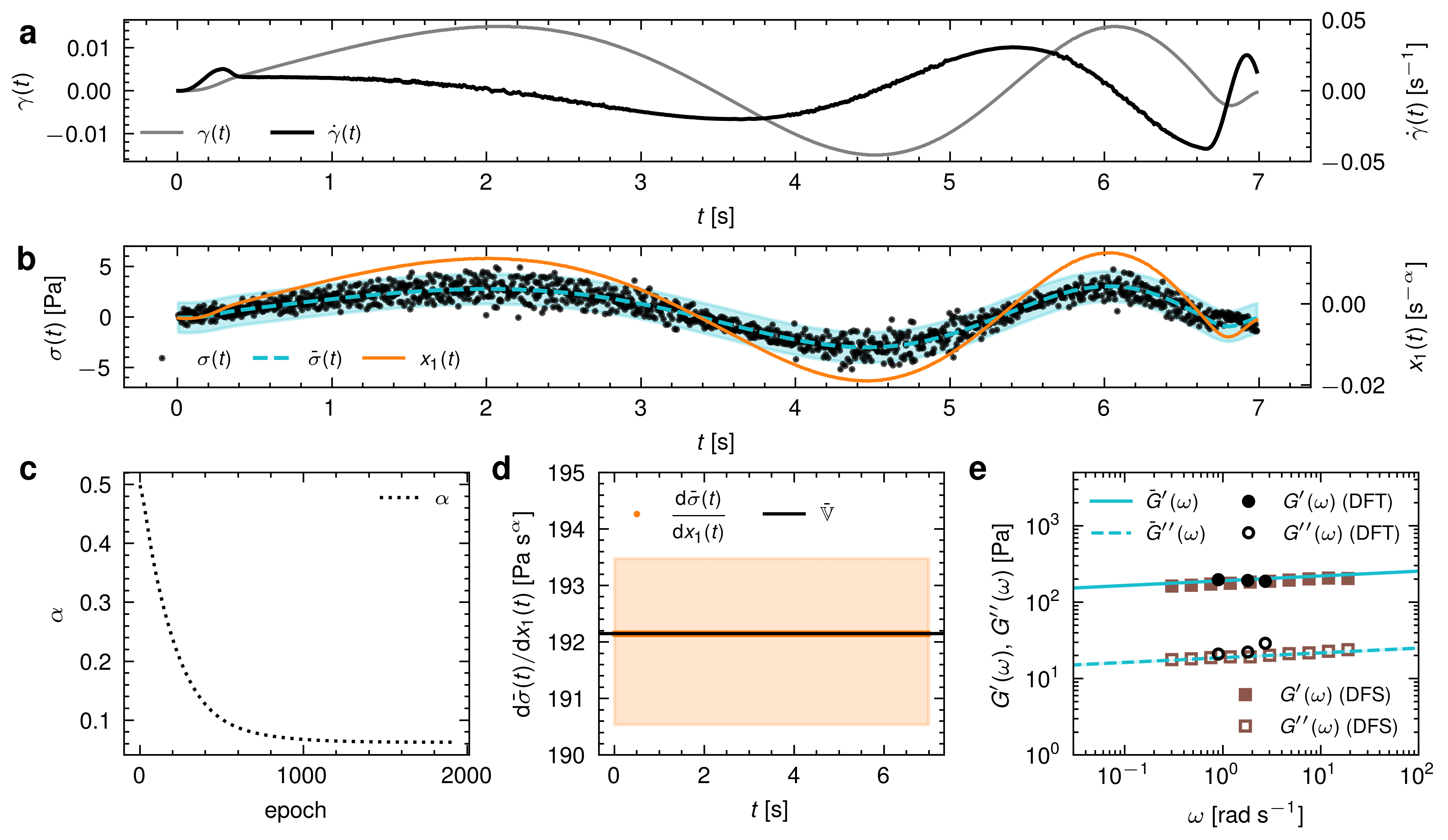}
    \caption{Time-domain, physics-informed Gaussian-process analysis of a chirp rheology experiment on a cellulose nanofiber hydrogel. \textbf{a} Chirp input in the time domain, showing the imposed strain $\gamma(t)$ on the left axis and strain rate $\dot{\gamma}(t)$ on the right axis. The instantaneous angular frequency spans the range $\omega \in [0.6,3]~\mathrm{rad\,s^{-1}}$. \textbf{b} GP fit to the measured stress $\sigma(t)$. Black markers denote the experimental data, the dashed curve is the posterior predictive mean $\bar{\sigma}(t)$, and the shaded region is the 95\% credible band. The engineered Scott~Blair memory feature $x_1(t)$ is shown on the secondary axis. \textbf{c} Training trajectory of the learned fractional exponent $\alpha$, demonstrating convergence. \textbf{d} Time-resolved sensitivity $\mathrm{d}\bar{\sigma}(t)/\mathrm{d}x_1(t)$ with pointwise credible bands. The horizontal line denotes the magnitude of the posterior mean quasi-property $\bar{\mathbb{V}}$. \textbf{e} Storage and loss moduli computed analytically using the parameters obtained from the learned time-domain model. Solid and dashed curves show $\bar{G}'(\omega)$ and $\bar{G}''(\omega)$, as predicted using the GP-inferred parameters. Circle markers show Discrete Fourier Transform (DFT) estimates from the same chirp, and square markers show independent discrete-frequency-sweep measurements.}
    \label{fig:cnf}
\end{figure*}

\subsection*{Inferring scale-free memory in a critical gel}

We next apply the same framework to a cellulose nanofiber hydrogel, a soft network whose response is expected to approach that of a critical gel.~\citep{miranda-valdez_sourroubille_makinen_puente-cordova_puisto_koivisto_alava_2024} Unlike a Maxwell fluid, a critical gel has no intrinsic relaxation time. Its storage and loss moduli follow parallel power laws in frequency, and the phase angle between stress and strain is approximately independent of frequency.~\citep{mours_winter_1994} In the time domain, this behavior corresponds to scale-free memory.

We first establish this character by conventional means.  A discrete frequency sweep (DFS) in the linear viscoelastic regime, probing one frequency at a time, returns $G^{\prime}(\omega)$ and $G^{\prime\prime}(\omega)$ that are parallel and power-law in frequency, with a nearly frequency-independent phase angle (Fig.~\ref{fig:cnf}e, squares)---the signature of a critical gel. Obtaining this spectrum required a measurement lasting $451~\mathrm{s}$.  The question we now ask is whether the same constitutive information is present in a chirp experiment lasting $7~\mathrm{s}$.

The hydrogel is probed using an OWCh that sweeps from $\omega_1=0.6~\mathrm{rad\,s^{-1}}$ to $\omega_2=3~\mathrm{rad\,s^{-1}}$ over a total duration of approximately $T_{\mathrm{owc}}=7~\mathrm{s}$ (Fig.~\ref{fig:cnf}a). These conditions are set due to instrumental limitations. The measured stress signal is substantially noisier than in the micellar experiment (Fig.~\ref{fig:cnf}b). While the micellar reference data was obtained using a flagship separate motor-transducer (SMT) system (ARES-G2, TA) capable of isolating pure material responses, the hydrogel characterization was intentionally subjected to the architectural and inertial constraints of a standard combined motor-transducer (CMT) rheometer (MCR302, Anton Paar). Crucially, this provides a stringent test of whether the proposed time-domain method can robustly recover a constitutive memory from noisy measurements.

For a critical gel, the natural candidate memory is the SB model,~\citep{koeller_1984, jaishankar_mckinley_2013}
\begin{equation}
    G(t) = \mathbb{V}\,\phi_1(t;\alpha) \qquad
    \phi_1(t;\alpha) = \frac{t^{-\alpha}}{\Gamma(1-\alpha)}
    \label{eq:scott_blair_kernel}
\end{equation}
where $\alpha\in(0,1)$ is the fractional exponent, $\mathbb{V}$ is a quasi-property with units $\mathrm{Pa\,s^\alpha}$, and $\Gamma(\cdot)$ is the Gamma function. The corresponding causal feature is
\begin{equation}
    x_1(t;\alpha) = \int_{0}^t
    \frac{1}{\Gamma(1-\alpha)}(t-t')^{-\alpha}\dot{\gamma}(t')\,
    \mathrm{d}t'
    \label{eq:scott_blair_feature}
\end{equation}

Once the GP with linear covariance kernel is trained, its posterior mean $\bar{\sigma}(t)$ follows the noisy stress trace without overfitting, yielding a coefficient of determination of $R^2=0.863$ (Fig.~\ref{fig:cnf}b). The GP infers a noise standard deviation of $0.75~\mathrm{Pa}$ against an output scale $s_f=1.60~\mathrm{Pa}$, giving a signal-to-noise ratio value of $\mathrm{SNR}\approx 2.15$. The GP assigns $82.2\%$ of the total signal variance to the SB constitutive feature and only $17.8\%$ to observation noise. Even in a relatively noisy measurement, the dominant stress variation is therefore captured by the physics-informed Scott~Blair feature.

The fractional exponent converges to $\alpha = 0.063$ (Fig.~\ref{fig:cnf}c), indicating a response close to an elastic solid with a finite scale-free dissipative component. The sensitivity $\mathrm{d}\bar{\sigma}(t)/\mathrm{d}x_{1}(t)$ (Fig.~\ref{fig:cnf}d) in this case defines the quasi-property of the SB model,
\[
    \bar{\mathbb{V}} = 192.2~\mathrm{Pa\,s}^{\alpha} \qquad
    95\%~\mathrm{CI} = [190.5,\,193.5]~\mathrm{Pa\,s}^{\alpha}
\]
The learned exponent and quasi-property determine the frequency-domain SB response through
\begin{equation}
    G^{\prime}(\omega) =
    \mathbb{V}\omega^\alpha\cos\!\left(\frac{\pi\alpha}{2}\right)
    \qquad
    G^{\prime\prime}(\omega) =
    \mathbb{V}\omega^\alpha\sin\!\left(\frac{\pi\alpha}{2}\right)
\end{equation}

The spectra inferred from the $7~\mathrm{s}$ chirp agree closely with the $451~\mathrm{s}$ discrete frequency sweep (Fig.~\ref{fig:cnf}e): the same constitutive information is recovered from a measurement two orders of magnitude shorter.  By contrast, DFT estimates from the same noisy chirp show noticeable scatter and resolve only a sparse set of frequency-domain points, so the gain comes not from the chirp protocol alone but from inferring the constitutive law directly in the time domain.

\begin{figure*}[t]
    \centering
    \includegraphics[width=\linewidth]{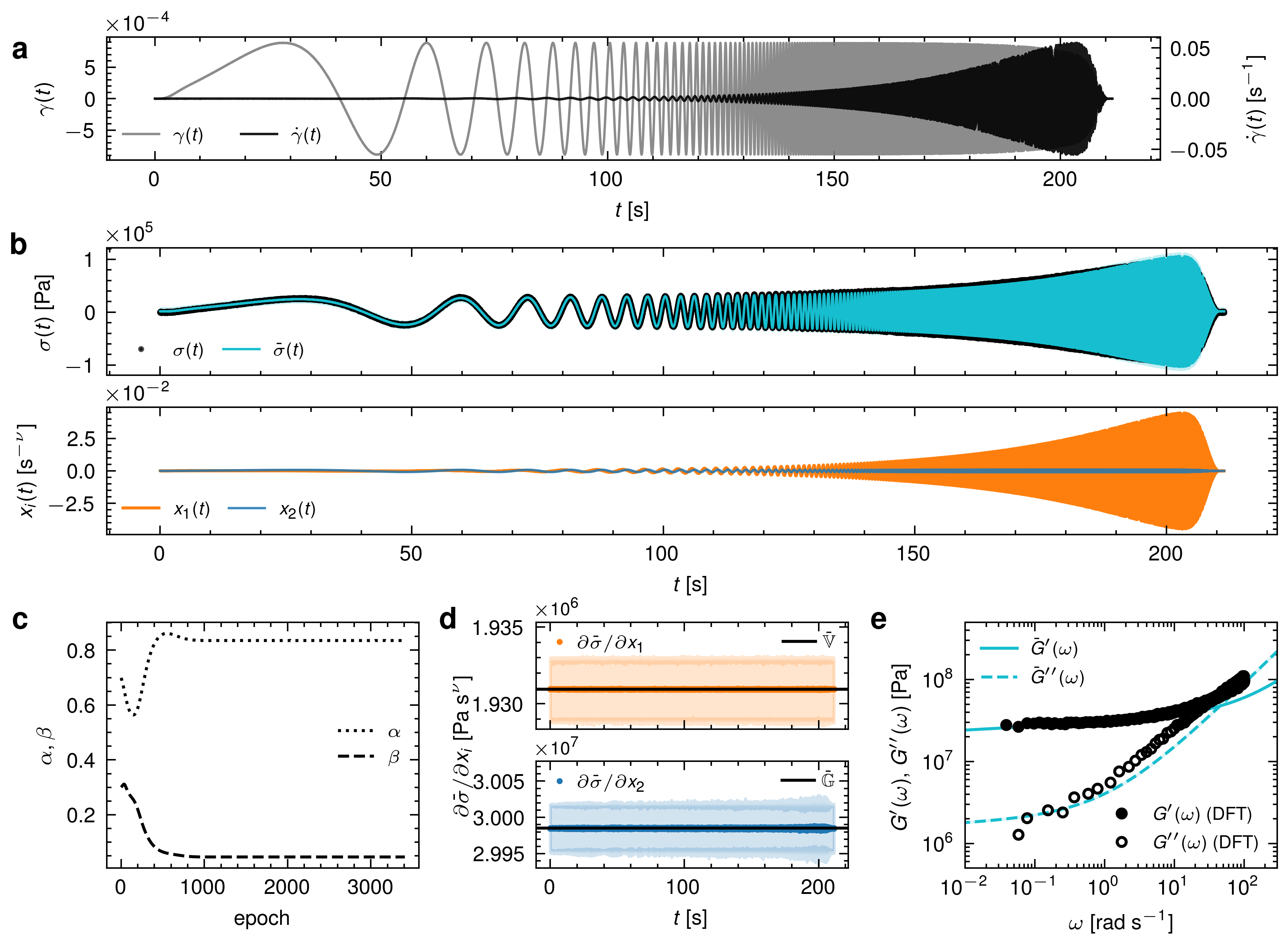}
    \caption{Time-domain, physics-informed Gaussian-process analysis of
    a chirp experiment on a photopolymer resin at a temperature $T=50~^\circ\mathrm{C}$. The experimental data corresponds to the data reported by~\citet{sheridan_zauscher_brinson_2024}.
    \textbf{a} Chirp input in the time domain, showing the imposed
    strain $\gamma(t)$ on the left axis and the strain rate
    $\dot{\gamma}(t)$ on the right axis; the instantaneous angular
    frequency spans $\omega\in[0.03,
    188.5]~\mathrm{rad\,s^{-1}}$.
    \textbf{b} Top: GP fit to the measured stress $\sigma(t)$. Black
    markers denote the experimental data, the solid curve is the
    posterior predictive mean $\bar{\sigma}(t)$, and the shaded region
    is the 95\% credible band, which is too tight and thus imperceptible in the figure. Bottom: the two causal memory features
    generated by the springpot branch, $x_1(t)$, and by the
    quasi-elastic branch, $x_2(t)$.
    \textbf{c} Training trajectories of the two fractional exponents,
    converging to $\alpha = 0.835$ and $\beta = 0.046$.
    \textbf{d} Time-resolved sensitivities of the predictive mean to
    each memory feature, with pointwise 95\% credible bands. The horizontal lines mark their posterior means,
    $\bar{\mathbb{V}}$ and $\bar{\mathbb{G}}$. The exponent $\nu$ in the
    axis label denotes $\alpha$ for $x_1$ and $\beta$ for $x_2$.
    \textbf{e} Storage and loss moduli evaluated from the learned
    time-domain model evaluated using the analytical solution of the FKV for the frequency domain. Solid and dashed cyan curves show
    $\bar{G}'(\omega)$ and $\bar{G}''(\omega)$ with uncertainty
    propagated from the joint posterior over both prefactors; filled and
    open markers show Discrete Fourier Transform (DFT) estimates from
    the same chirp.}
    \label{fig:fkv}
\end{figure*}

\subsection*{Apportioning the stress between two memory features}

In the previous examples, the stress was explained by a single memory feature, so the GP had only to scale one signal. A constitutive model with two relaxation mechanisms acting in parallel generates two features whose weighted sum is the measured stress; the question becomes whether one stress trace contains enough information to apportion the response between the two contributions---and with what confidence.

We test this on a photopolymer resin, in which a permanent crosslinked matrix is expected to coexist with a broadly distributed dissipative fraction. The chirp data are those reported by~\citet{sheridan_zauscher_brinson_2024} (Fig.~\ref{fig:fkv}a,b). The model family that describes this material is the Fractional Kelvin--Voigt (FKV) model.~\citep{miranda-valdez_niinisto_makinen_lejon_koivisto_alava_2025, song_holten-andersen_mckinley_2023} The full FKV consists of two Scott~Blair (springpot) elements in parallel,
\begin{equation}
    G(t) = \frac{\mathbb{V}}{\Gamma(1-\alpha)}\,t^{-\alpha}
         + \frac{\mathbb{G}}{\Gamma(1-\beta)}\,t^{-\beta}
    \qquad t>0
    \label{eq:fkv_relax_results}
\end{equation}
with quasi-properties $\mathbb{V}$ [$\mathrm{Pa\,s}^{\alpha}$] and $\mathbb{G}$ [$\mathrm{Pa\,s}^{\beta}$]. Its limits carry distinct physics: $\beta=0$ reduces the second term of Eq.~\eqref{eq:fkv_relax_results} to a Hookean spring (FKV-S), while $\alpha=1$ reduces the first to a Newtonian dashpot (FKV-D); retaining one term recovers the single springpot (SB). Because the branches act in parallel their stresses add, so Eq.~\eqref{eq:features} generates one causal feature per branch,
\begin{equation}
    \sigma(t) \;\approx\; \mathbb{V}\,x_1(t;\alpha)
                        + \mathbb{G}\,x_2(t;\beta)
    \label{eq:fkv_superposition}
\end{equation}

To train the GP with linear covariance kernel, feature construction is decoupled---each branch filters the same strain-rate history through its own kernel---while inference is coupled, since both features derive from one $\dot\gamma(t)$ and are partially collinear. The GP therefore returns a joint posterior over $(\mathbb{V},\mathbb{G})$ whose width reflects how well the measurement separates the two mechanisms.

\begin{table*}[t]
  \centering
  \caption{Model comparison for the photopolymer resin chirp ($N=32{,}689$). The fit term $2\mathcal{U}$ and the complexity penalties $2k$ and $k\ln N$ are listed separately, so that the origin of each ranking is visible; $k$ counts memory-shape parameters, prefactors, and GP hyperparameters. Only differences between candidates fitted to the same record are meaningful; absolute values are given for reference. Lower $\Delta$ is better; $\Delta = 0$ marks the selected model. Dashes denote parameters absent from the model. The prefactors $c_1$ and $c_2$ are the constitutive quantities recovered as GP sensitivities, and differ between models: $(\mathbb{V},\mathbb{G})$ for the FKV, in $\mathrm{Pa\,s}^{\alpha}$ and $\mathrm{Pa\,s}^{\beta}$; $(\mathbb{V},G_0)$ for the FKV-S, in $\mathrm{Pa\,s}^{\alpha}$ and Pa; $(\eta,\mathbb{G})$ for the FKV-D, in $\mathrm{Pa\,s}$ and $\mathrm{Pa\,s}^{\beta}$; and $\mathbb{V}$ alone for the SB model, in $\mathrm{Pa\,s}^{\alpha}$.}
  \label{tab:fkv_ic}
  \small
  \setlength{\tabcolsep}{3pt}
  \renewcommand{\arraystretch}{1.2}
  \begin{tabular}{lcccccccccccc}
    \hline
    Model & $2\mathcal{U}$ & $2k$ & $k\ln N$ & AIC & BIC & $\Delta$AIC & $\Delta$BIC & RMSE [Pa] & $c_1$ & $c_2$ & $\alpha$ & $\beta$ \\
    \hline
FKV   & $-73134.4$ & 16 & 83.2 & $-73118.4$ & $-73051.2$ & 0                  & 0                  & 2439  & $1.93\times10^{6}$ & $3.00\times10^{7}$ & 0.835 & 0.046 \\
    FKV-S & $-67272.5$ & 14 & 72.8 & $-67258.5$ & $-67199.7$ & $5.86\times10^{3}$ & $5.85\times10^{3}$ & 2671  & $2.94\times10^{6}$ & $2.87\times10^{7}$ & 0.756 & --    \\
    FKV-D & $-62536.9$ & 14 & 72.8 & $-62522.9$ & $-62464.1$ & $1.06\times10^{4}$ & $1.06\times10^{4}$ & 2875  & $8.54\times10^{5}$ & $3.16\times10^{7}$ & --    & 0.097 \\
    SB    & $25003.5$  & 12 & 62.4 & $25015.5$  & $25065.8$  & $9.81\times10^{4}$ & $9.81\times10^{4}$ & 11037 & $1.81\times10^{7}$ & --                 & 0.382 & --    \\
    \hline
  \end{tabular}
\end{table*}

Based on the available chirp data, both criteria select the full FKV model (Table~\ref{tab:fkv_ic}). The residuals of the selected model are correlated over $\tau_{\mathrm{int}}=3.2$ samples, so the $N$ time samples are effectively independent and the assumption underlying Eq.~\eqref{eq:bic} is satisfied. Recomputing the criteria at the corresponding effective sample size for every model in Table~\ref{tab:fkv_ic} leaves the ordering unchanged.

The selected model, FKV, places the second branch close to the elastic limit, $\beta=0.046$, and the data resolve this small exponent decisively: fixing $\beta=0$ (FKV-S) costs $\Delta\mathrm{BIC}\approx5.9\times10^{3}$, of which only $10.4$ is the penalty for the additional parameter, the remainder being a genuine loss of fit that also raises the RMSE by $10\%$ ($2671$ vs.\ $2439$~Pa). The constraint additionally distorts the apportionment: forced to represent the dissipative branch without a compensating power law in the network, the FKV-S inflates $\mathbb{V}$ by $52\%$ ($2.94$ vs.\ $1.93\times10^{6}~\mathrm{Pa\,s}^{\alpha}$) while leaving the network prefactor within $5\%$.

The distinction is qualitative in frequency space. The loss contribution of a springpot of order $\beta$ scales as $\mathbb{G}\omega^{\beta}\sin(\pi\beta/2)$, so the branch is entirely absent from $G''(\omega)$ when $\beta=0$ and the loss modulus reduces to the single power law $\mathbb{V}\omega^{\alpha}\sin(\pi\alpha/2)$ of the dissipative branch. A nonzero $\beta$ activates a nearly frequency-independent contribution of $\mathbb{G}\sin(\pi\beta/2)\approx2.2\times10^{6}~\mathrm{Pa}$, which is negligible against the dissipative branch at high frequency ($3\%$ at $\omega=100~\mathrm{rad\,s^{-1}}$) but dominates it below $\omega\approx1.2~\mathrm{rad\,s^{-1}}$, exceeding it nineteenfold at the bottom of the excited band. The evidence therefore resolves a weak power-law relaxation of the network branch, which turns $G''(\omega)$ from a single power law into a quasi-plateau at low frequency.

The posterior mean tracks the stress with $R^2=0.994$ (Fig.~\ref{fig:fkv}b, top). The two features beneath it weight the same history differently---$x_1$ follows the strain rate and grows toward high frequency, $x_2$ is essentially the strain itself as $\beta \to 0$---and this difference in phase and envelope is what lets the GP separate them. Both exponents converge within a few thousand epochs and remain flat (Fig.~\ref{fig:fkv}c). The sensitivities (Fig.~\ref{fig:fkv}d) return the quasi-property values
\[
\bar{\mathbb{V}} = 1.93\times10^{6}~\mathrm{Pa\,s}^{\alpha}
\qquad
\bar{\mathbb{G}} = 3.00\times10^{7}~\mathrm{Pa\,s}^{\beta}
\]

Evaluating the FKV material functions (explicit forms in Supplementary Information, Sec. S4) with the inferred parameters, reproduces the quasi-elastic plateau at low frequency in $G^{\prime}(\omega)$ and the power-law rise of both moduli at high frequency (Fig.~\ref{fig:fkv}e). $G'$ agrees with the DFT estimates across the band; $G''$ agrees where dissipation is appreciable but the DFT estimates exceed the model at low frequency, where $G''/G'\lesssim10^{-2}$ makes the out-of-phase component a small difference of large signals, dominated by leakage and noise. The constitutive inference regularizes exactly this weakly determined quantity at low frequencies.

\subsection*{Detecting non-stationarity in a mutating material}

\begin{figure}[t]
    \centering
    \includegraphics[width=\linewidth]{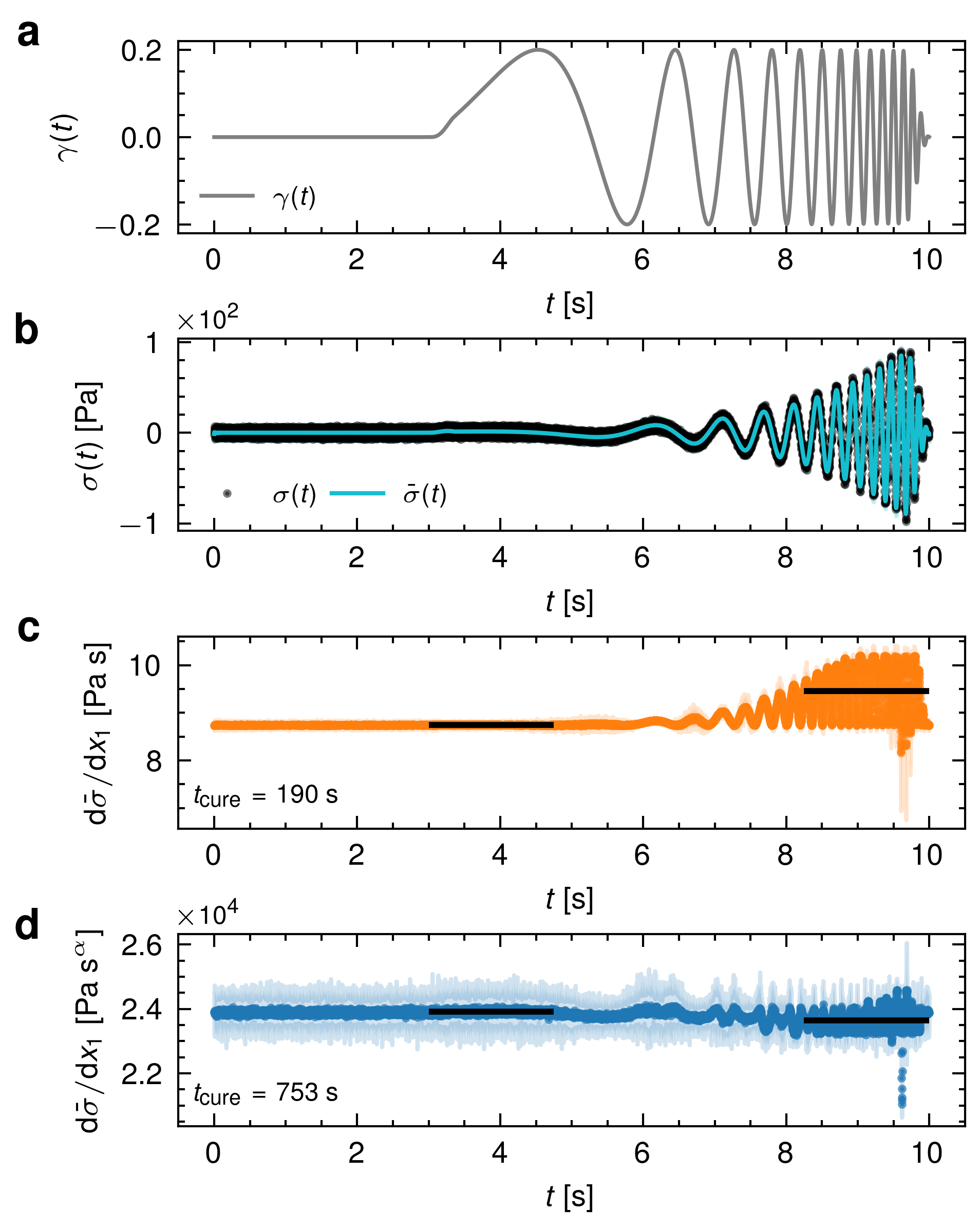}
    \caption{Time-domain, physics-informed Gaussian-process analysis of a UV-curable acrylate system from \citet{perego_vadillo_mills_das_mckinleyfrs_2025}, measured at $\approx$190~s after the onset of UV irradiation. \textbf{a} Chirp input $\gamma(t)$; the instantaneous angular frequency spans the range $\omega\in[0.6,60]~\mathrm{rad\,s^{-1}}$. \textbf{b} GP fit to the measured stress $\sigma(t)$. Black markers denote the experimental data, the dashed curve is the posterior predictive mean $\bar{\sigma}(t)$, and the shaded region is the 95\% credible band, which is too tight and imperceptible in the figure. \textbf{c} Time-resolved sensitivity $\mathrm{d}\bar{\sigma}(t)/\mathrm{d}x_1(t)$ of the GP prediction to the Scott~Blair memory feature. \textbf{d} Time-resolved sensitivity $\mathrm{d}\bar{\sigma}(t)/\mathrm{d}x_1(t)$ of the GP prediction to the Scott~Blair memory feature for the same acrylate system but at a later curing stage. Solid black lines in panels \textbf{c} and \textbf{d} represent the averaged sensitivity at the start and end of each chirp.}
    \label{fig:acrylic}
\end{figure}

The preceding sections demonstrate the framework for time-invariant materials whose constitutive memory does not change during the measurement. We now ask whether the GP sensitivity can detect active material mutation during the chirp itself---in other words, a direct violation of the time-translation-invariant (TTI) principle that underpins the Boltzmann superposition integral.

We apply the GP framework to the chirp data measured by \citet{perego_vadillo_mills_das_mckinleyfrs_2025} for a UV-curable acrylate system. These authors measured the material response at $t_{\mathrm{cure}}\approx190~\mathrm{s}$ after the onset of UV irradiation, where they report the material mutation to be highest, violating the TTI principle. The material is modeled with a GP with RBF kernel using the Scott~Blair convolutional feature from Eq.~\eqref{eq:scott_blair_feature}. The optimized fractional exponent converges to $\alpha=0.9999\approx1$, which collapses the springpot to a Newtonian dashpot and assigns the quasi-property $\mathbb{V}$ units of dynamic viscosity, $\eta$. This limiting behavior is physically consistent with a partially crosslinked acrylate network in which viscous dissipation still dominates over elastic storage at the probed frequencies and curing stage.~\citep{perego_vadillo_mills_das_mckinleyfrs_2025}

The GP posterior mean reproduces the measured stress with $R^2=0.950$ (Fig.~\ref{fig:acrylic}a,b). The inferred noise standard deviation is $4.30~\mathrm{Pa}$ against an output scale of $s_f=42.5~\mathrm{Pa}$, giving a signal-to-noise ratio of $\approx 9.9$ and attributing $99.0\%$ of the total stress variance to the constitutive Scott~Blair feature rather than to observation noise.

Because $\alpha\to1$, the Scott~Blair feature reduces to the instantaneous strain rate, $x_1(t)\simeq\dot{\gamma}(t)$, so the sensitivity $\mathrm{d}\bar{\sigma}(t)/\mathrm{d}x_1(t)$ acquires the units and meaning of a dynamic viscosity, $\eta(t)$. Over the first three seconds the sample is held at zero strain: the feature is a linear functional of $\dot{\gamma}$, so it vanishes identically, and every time point in that interval maps to the same location at the origin of the stress--feature space. The value returned there, $\mathrm{d}\bar{\sigma}/\mathrm{d}x_1\approx8.8~\mathrm{Pa\,s}$, is therefore the slope of the learned map rather than a data-constrained quantity, and we exclude it from the averages below. Once the chirp begins, the sensitivity is constrained by the data. It rises steadily and develops oscillations of growing amplitude, approaching $10.2~\mathrm{Pa\,s}$ by the end of the sweep (Fig.~\ref{fig:acrylic}c). Averaged over the first and last quarters of the sweep, it increases by $+8.1\%$. The magnitude agrees with \citet{perego_vadillo_mills_das_mckinleyfrs_2025}, who report a transition from a plateau to a peak complex viscosity near $10~\mathrm{Pa\,s}$ for the same dataset from the Discrete Fourier Transform of the measured stress. The growth of $\eta(t)$ over a single chirp is thus resolved directly in the time domain, without a prescribed cure kinetics or prior knowledge of the crosslinking dynamics.

The excitation amplitude of $\dot{\gamma}$ also grows through the sweep, so later times correspond to larger feature magnitudes, and a drifting sensitivity could in principle reflect that growth rather than a change in the material. A control measurement addresses this. Applying the same RBF-kernel analysis to a chirp recorded on the same sample at $t_{\mathrm{cure}}=753~\mathrm{s}$, where the mutation number is no longer appreciable, gives a sensitivity that varies by $-1.1\%$ between the same two windows (Fig.~\ref{fig:acrylic}d), compared with $+8.1\%$ for the record at $190~\mathrm{s}$. The capacity of the GP to represent a nonlinear stress--feature map is governed by the ratio of the feature range to the RBF length scale $\ell$. For the control fit this ratio is $8.1/2.24 = 3.6$, and for the mutating fit $9.2/4.29 = 2.1$ (higher ratios indicate greater kernel nonlinearity), both in standardized feature units. The control therefore admits more curvature over its feature range than the mutating fit, and returns a constant sensitivity. The drift observed at $190~\mathrm{s}$ is therefore attributable to material evolution rather than to growth of the excitation, and constitutes a violation of time-translation invariance detected within a single chirp.

The fitted exponents differ between the two records. At $t_{\mathrm{cure}}=190~\mathrm{s}$, $\alpha=0.9999$, so the response is effectively Newtonian and the sensitivity has units of viscosity; at $t_{\mathrm{cure}}=753~\mathrm{s}$, $\alpha=0.721$, indicating a fractional response with appreciable elastic storage. The quasi-properties of the two fits accordingly carry different units, $\mathrm{Pa\,s}$ and $\mathrm{Pa\,s}^{\alpha}$, and their absolute magnitudes are not directly comparable; the quantity compared above is the relative change within each record. The decrease in $\alpha$ between the two measurements provides an independent indication of the crosslinking that the time-resolved sensitivity resolves within the earlier record.

\section*{Discussion}

We have introduced a physics-informed GP framework for inferring viscoelastic constitutive behavior directly from the time-domain response of a chirp experiment. The approach exploits continuously time-resolved data in four ways.

First, rather than compressing the measured time series into frequency-domain moduli and fitting a model to those estimates, we treat the complete stress--strain record as the object of inference. Frequency-domain moduli are recovered analytically from the learned memory kernel, with uncertainty propagated from the time-domain GP model. This is particularly valuable for short or noisy records---exactly the conditions where chirp rheometry is most attractive---where Fourier estimates are sparse and affected by finite-window artifacts. The CNF hydrogel example illustrates this directly: the GP recovers a well-defined Scott~Blair exponent and quasi-property $\mathbb{V}$ from a noisy chirp that defeats conventional DFT analysis, attributing $82.2\%$ of the total stress variance to the constitutive feature rather than to noise. 

Second, unlike black-box regressors, the framework returns directly interpretable rheological quantities: relaxation times, fractional exponents, moduli, and quasi-properties, the latter two with quantified uncertainty. With a linear GP covariance kernel, the framework is equivalent to Bayesian linear regression in the physics-informed feature space, and the feature sensitivities are the constitutive prefactors that scale the stress--strain relationship linearly. With an RBF kernel, the same sensitivities become local slopes of the learned stress--feature map. Their time-constancy certifies linear viscoelastic consistency, while systematic variation flags phenomena such as material mutation during the measurement. 

It is worth stating explicitly what the probabilistic formulation contributes beyond a direct nonlinear least-squares fit of Eq.~\eqref{eq:features}. First, the prefactors $c_i$ enter linearly and are marginalized analytically rather than optimized, so model comparison carries the corresponding Occam factor instead of treating each prefactor as a free parameter fitted only through residual minimization. Second, the observation-noise variance is inferred jointly with the constitutive parameters rather than assumed, enabling the variance decompositions reported for the CNF hydrogel and the acrylate system. Third, the covariance structure is itself a modeling choice: replacing the linear kernel with an RBF kernel converts the framework from a parametric estimator of constitutive prefactors into a nonparametric diagnostic of whether a linear, time-translation-invariant constitutive relation holds at all. This last capability has no direct counterpart in a least-squares fit of a prescribed constitutive form.

Third, the result in Fig.~\ref{fig:chirp_90} demonstrates that a physics-informed GP trained on one oscillatory record correctly predicts the response to an independent record with a different bandwidth, duration, and initial phase. This generalization is guaranteed by construction once $G(t;\bm\Theta)$ is identified, but it is not automatic: if the inferred kernel were merely a compressed description of the training waveform, the out-of-protocol prediction would fail, and it would fail first in the frequency range that the training record did not excite. The agreement at the terminal crossover, an order of magnitude below the training band, confirms that the GP has internalized a genuine constitutive law rather than a waveform interpolator. This capability is not achieved by other machine-learning methods reported in the literature, such as rheology-informed neural networks.~\citep{saadat_mahmoudabadbozchelou_jamali_2022}

Fourth, the UV-curable acrylate example demonstrates that the time-resolved GP sensitivity provides a model-free probe of material mutation. For the acrylate system, when $\alpha\to1$ and the Scott~Blair feature reduces to the instantaneous strain rate, the sensitivity becomes directly proportional to the instantaneous dynamic viscosity $\eta(t)$. The monotonic rise of the sensitivity after $t\approx3~\mathrm{s}$ in Fig.~\ref{fig:acrylic}c is therefore a direct measurement of the viscosity increase driven by UV-induced crosslinking, extracted without any prescribed kinetic model. More generally, TTI violations that produce systematic changes in the local stress--feature relation can manifest as sensitivity drift, providing a quantitative and protocol-independent fingerprint of structural evolution.

The present work is restricted to the linear viscoelastic regime. Natural extensions include discrete relaxation kernels (e.g. Generalized Maxwell model), nonlinear extensions coupling memory to strain amplitude for LAOS-like protocols, and adaptive chirp design that optimizes the signal to discriminate between competing constitutive models. More broadly, the results establish physics-informed feature construction as a route to generalizable machine learning in rheology. The method is not tied to a particular waveform: any deformation history can be transformed into causal memory features once a candidate relaxation kernel is specified, enabling model libraries, uncertainty-aware parameter estimation, and evidence-based selection among competing physical hypotheses---all from the raw output of a single short time-resolved experiment.

\section*{Materials and methods}

The chirp rheometry protocols, physics-informed feature construction, Gaussian-process inference framework (including covariance kernels, variational training, and sensitivity-based prefactor recovery), frequency-domain reconstruction procedure, and model-selection methodology are described in full in the Supplementary Information (Secs.~S1--S6). All features and stresses are standardized to zero mean and unit variance before training; parameters retain physical units after inverting the scaling. Parameter constraints ($0\le\beta<\alpha\le1$; positive relaxation times) are enforced via smooth reparameterizations during optimization. The framework is implemented in Python using GPyTorch~\citep{gardner2018gpytorch}; code and data are openly available at \url{https://github.com/mirandi1/rheogp}. The repository was developed with assistance from Claude Sonnet 4.6 (Anthropic).

\section*{Supplementary Materials}
See the Supplementary Materials for the chirp rheometry protocols and experimental datasets (Sec.~S1), the construction and numerical evaluation of the physics-informed memory features (Sec.~S2), the Gaussian-process inference framework, covariance kernels, variational training and model-selection methodology (Sec.~S3), the complete model library with constitutive equations, material functions and the frequency-domain reconstruction procedure (Sec.~S4), the residual diagnostics and the assessment of the independence assumption (Sec.~S5), and the implementation details (Sec.~S6). Tables S1 and S2 are included.

\section*{References}
\bibliography{article_main}

\begin{acknowledgments}
\textbf{Funding:} I.~M.~V. thanks the Vilho, Yrj\"{o}, and Kalle V\"{a}is\"{a}l\"{a} Foundation of the Finnish Academy of Science and Letters and the Finnish Foundation for Technology Promotion for personal funding. The authors acknowledge funding from the FinnCERES flagship (151830423), Business Finland (210129, 211835, 211909, 211989), the Future Makers program, and the Finnish Cultural Foundation. The authors acknowledge funding from the European Union: the ARCHIBIOFOAM project has received funding from the European Union's Horizon Europe research and innovation programme under grant agreement No 101161052. Views and opinions expressed are, however, those of the authors only and do not necessarily reflect those of the European Union or the European Innovation Council and SMEs Executive Agency (EISMEA). Neither the European Union nor the granting authority can be held responsible for them. The authors acknowledge the computational resources provided by the Aalto Science-IT project, and thank Mohua Das and Prof.\ Gareth McKinley for kindly sharing data and for their guidance on the numerical simulation of chirps.

\textbf{Author contributions:} I.~M.~V.: Conceptualization, Methodology, Software, Formal analysis, Visualization, Investigation, Writing---original draft, Writing---review and editing. J.~K.: Supervision, Validation, Writing---review and editing. M.~A.: Supervision, Validation, Writing---review and editing, Funding acquisition, Project administration.

\textbf{Competing interests:} The authors declare that they have no competing interests.

\textbf{Data and materials availability:} All data needed to evaluate the conclusions in the paper are present in the paper and/or the Supplementary Materials. The data and code that support the findings of this study are openly available at \url{https://github.com/mirandi1/rheogp}.
\end{acknowledgments}

\end{document}


\title{Supplementary Information: Direct inference of viscoelastic memory from chirp rheometry via physics-informed Gaussian processes}

\author{Isaac Y. Miranda-Valdez}
\email[]{isaac.miranda.valdez@alumni.aalto.fi}
\affiliation{Department of Applied Physics, School of Science, Aalto University, P.O.\ Box 15600, Espoo, FI-00076 Aalto, Finland}
\affiliation{Deltan Labs, P.O.\ Box 11000, Espoo, FI-00076 Aalto, Finland}

\author{Juha Koivisto}
\affiliation{Department of Applied Physics, School of Science, Aalto University, P.O.\ Box 15600, Espoo, FI-00076 Aalto, Finland}

\author{Mikko J. Alava}
\email[]{mikko.alava@aalto.fi}
\affiliation{Department of Applied Physics, School of Science, Aalto University, P.O.\ Box 15600, Espoo, FI-00076 Aalto, Finland}

\date{\today}
\maketitle

\section*{S1.\; Chirp rheometry protocol and datasets}
\label{sec:SI_chirp}

All time-domain data, except those for the photopolymer resin, were obtained from strain-controlled optimally windowed chirp ($\gamma$-OWCh) protocols following Geri \textit{et al.}~\cite{geri_keshavarz_divoux_clasen_curtis_mckinley_2018}. The imposed strain is
\begin{equation}
    \gamma(t) = \gamma_0\,w(t)\,\sin[\psi(t)],
    \label{eq:SI_chirp}
\end{equation}
where $\gamma_0$ is the nominal strain amplitude, $w(t)$ is a dimensionless tapering window, and $\psi(t)$ is the accumulated phase. To probe the frequency interval $[\omega_1,\omega_2]$ over a duration $T_{\mathrm{owc}}$, the chirps use a logarithmic sweep,
\begin{equation}
    \psi(t) = \frac{\omega_1 T_{\mathrm{owc}}}{\ln R}\left[\exp\!\left(\frac{t}{T_{\mathrm{owc}}}\ln R\right) - 1\right] + \psi_0, \qquad R = \frac{\omega_2}{\omega_1},
    \label{eq:SI_log_sweep}
\end{equation}
so that the instantaneous angular frequency $\omega(t) = \mathrm{d}\psi/\mathrm{d}t$ increases exponentially from $\omega_1$ to $\omega_2$; $\psi_0$ is an initial phase offset, zero unless stated otherwise. To reduce spectral leakage, a chirp is tapered with a Tukey-type window,
\begin{equation}
    w(t) =
    \begin{cases}
    \cos^2\!\left[\dfrac{\pi}{r}\!\left(\dfrac{t}{T_{\mathrm{owc}}}-\dfrac{r}{2}\right)\right], & 0\leq \dfrac{t}{T_{\mathrm{owc}}}\leq \dfrac{r}{2},\\[8pt]
    1, & \dfrac{r}{2}<\dfrac{t}{T_{\mathrm{owc}}}<1-\dfrac{r}{2},\\[8pt]
    \cos^2\!\left[\dfrac{\pi}{r}\!\left(\dfrac{t}{T_{\mathrm{owc}}}-1+\dfrac{r}{2}\right)\right], & 1-\dfrac{r}{2}\leq \dfrac{t}{T_{\mathrm{owc}}}\leq 1,
    \end{cases}
    \label{eq:SI_tukey}
\end{equation}
where $r\in(0,1)$ is the taper fraction; the reported data used $r=0.10$ throughout, except for the photopolymer resin. Because the taper drives the strain to zero at both ends of the record, the measured signal may be treated as one period of a periodic protocol, a property exploited by the frequency-domain evaluation of the memory features (Sec.~S2.1).

Several records begin with a quiescent interval during which the imposed strain is held at zero before the sweep starts. In that interval the memory features of Eq.~\eqref{eq:SI_features} vanish identically, so the measured stress is instrument noise alone; this is used in Sec.~S5 to separate the noise autocorrelation of the instrument from the residual autocorrelation of the fitted models.

\subsection*{Experimental datasets}

\textit{Wormlike micellar solution (training record).} We analyze the dataset reported by Das \textit{et al.}~\cite{das_vadillo_perego_mckinley_2026}. The chirp spans $\omega_1=3~\mathrm{rad\,s^{-1}}$ to $\omega_2=30~\mathrm{rad\,s^{-1}}$ over $T_{\mathrm{owc}}=2\pi/\omega_1\approx2.0~\mathrm{s}$, with strain amplitude $\gamma_0=0.15$, $r=0.10$ and $\psi_0=0$, sampled at $500~\mathrm{Hz}$, and preceded by a $1~\mathrm{s}$ quiescent interval at zero strain. The complete record contains $N=1499$ samples. The measurement was performed at $T=25~^\circ\mathrm{C}$ on a separate motor--transducer (SMT) instrument (ARES-G2, TA Instruments). Consult the work of \citet{das_vadillo_perego_mckinley_2026} for further details.

\textit{Wormlike micellar solution (out-of-protocol test record).} The protocol-independence test of the main text (Fig.~2) evaluates the GP trained on the record above on an independent chirp measured on the same material. This second chirp spans $\omega_1=0.3~\mathrm{rad\,s^{-1}}$ to $\omega_2=30~\mathrm{rad\,s^{-1}}$ over $T_{\mathrm{owc}}=14~\mathrm{s}$, and begins with an initial phase offset $\psi_0=-90^{\circ}$, a feature absent from the training record. It is likewise preceded by a $1~\mathrm{s}$ quiescent interval at zero strain. Consult the work of \citet{das_vadillo_perego_mckinley_2026} for further details.

\textit{Cellulose nanofiber (CNF) hydrogel.} The chirp spans $\omega_1=0.6~\mathrm{rad\,s^{-1}}$ to $\omega_2=3~\mathrm{rad\,s^{-1}}$ over $T_{\mathrm{owc}}\approx7~\mathrm{s}$, with $\gamma_0=0.015$ (verified in the linear viscoelastic regime by independent amplitude-sweep measurements), $r=0.10$ and $\psi_0=0$, sampled at $200~\mathrm{Hz}$. The complete record contains $N=1349$ samples. The experiment was performed at $T=10~^\circ\mathrm{C}$ in an Anton Paar MCR 302 rheometer, a combined motor--transducer (CMT) instrument, with a profiled parallel-plate geometry (PP25). The band and duration were set by instrumental limitations. The independent discrete frequency sweep (DFS) used for comparison in the main text was measured on the same sample, one frequency at a time, over a total duration of $451~\mathrm{s}$.

\textit{Photopolymer resin.} The chirp data are taken from \citet{sheridan_zauscher_brinson_2024}. The record contains $N=32{,}689$ samples over ${\sim}210~\mathrm{s}$ at a strain amplitude $\gamma_0\sim10^{-4}$. These authors used a chirp protocol different from the one established by \citet{geri_keshavarz_divoux_clasen_curtis_mckinley_2018} owing to instrumental limitations. The chirp spans $\omega_1=0.03~\mathrm{rad\,s^{-1}}$ to $\omega_2=188.5~\mathrm{rad\,s^{-1}}$, with a tapering parameter $r=0.05$ and an effective sampling frequency of $156.78~\mathrm{Hz}$. The measurement was performed at $T=50~^\circ\mathrm{C}$. Consult the work of \citet{sheridan_zauscher_brinson_2024} for further details.

\textit{UV-curable acrylate (mutating record).} The chirp data are taken from \citet{perego_vadillo_mills_das_mckinleyfrs_2025}, measured at $t_{\mathrm{cure}}\approx190~\mathrm{s}$ after the onset of UV irradiation, the stage at which those authors report the mutation number to be highest. The chirp spans $\omega_1=0.6~\mathrm{rad\,s^{-1}}$ to $\omega_2=60~\mathrm{rad\,s^{-1}}$ over $T_{\mathrm{owc}}=4\pi/(3\omega_1)\approx7~\mathrm{s}$, with strain amplitude $\gamma_0=0.20$, $r=0.10$ and $\psi_0=0$, sampled at $500~\mathrm{Hz}$, and preceded by a ${\sim}3~\mathrm{s}$ pre-conditioning interval at zero strain. Consult the work of \citet{perego_vadillo_mills_das_mckinleyfrs_2025} for further details.

\textit{UV-curable acrylate (stationary control record).} The control measurement of the main text (Fig.~5d) is a chirp recorded on the same sample at $t_{\mathrm{cure}}=753~\mathrm{s}$, by which stage the mutation number reported by \citet{perego_vadillo_mills_das_mckinleyfrs_2025} is no longer appreciable. The chirp protocol, taper fraction, and sampling rate are those of the mutating record, so that the two differ only in cure time. Consult the work of \citet{perego_vadillo_mills_das_mckinleyfrs_2025} for further details.

\section*{S2.\; Physics-informed memory features}
\label{sec:SI_features}

The causal convolutional features used as inputs to the Gaussian process are constructed from the linear viscoelastic Boltzmann superposition principle. For an imposed strain-rate history $\dot{\gamma}(t)$,
\begin{equation}
    \sigma(t) = \int_{0}^{t} G(t-t')\,\dot{\gamma}(t')\,\mathrm{d}t',
    \label{eq:SI_boltzmann}
\end{equation}
where $\sigma(t)$ is the shear stress and $G(t)$ is the relaxation modulus. Writing the candidate relaxation modulus as
\begin{equation}
    G(t;\boldsymbol{\Theta}) = \sum_{i=1}^{n} c_i\,\phi_i(t;\boldsymbol{\kappa}), \qquad t>0,
    \label{eq:SI_kernel_decomp}
\end{equation}
and substituting into Eq.~\eqref{eq:SI_boltzmann} gives
\begin{equation}
    \sigma(t) = \sum_{i=1}^{n} c_i\,x_i(t;\boldsymbol{\kappa}),
    \label{eq:SI_feature_stress}
\end{equation}
where the causal convolutional features are
\begin{equation}
    x_i(t;\boldsymbol{\kappa}) = \int_{0}^{t} \phi_i(t-t';\boldsymbol{\kappa})\,\dot{\gamma}(t')\,\mathrm{d}t', \qquad i=1,\ldots,n.
    \label{eq:SI_features}
\end{equation}
Here $\boldsymbol{\Theta}=(\boldsymbol{\kappa},\mathbf{c})$ denotes the full parameter set, $\boldsymbol{\kappa}$ contains the memory-shape parameters (relaxation times, fractional exponents), and $\mathbf{c}=(c_1,\ldots,c_n)^\top$ contains the prefactors (moduli, viscosities, quasi-properties). The GP input at each time point is the feature vector
\begin{equation}
    \mathbf{x}(t;\boldsymbol{\kappa}) = \bigl[x_1(t;\boldsymbol{\kappa}),\;\ldots,\; x_n(t;\boldsymbol{\kappa})\bigr]^\top \in\mathbb{R}^{n}.
    \label{eq:SI_feature_vector}
\end{equation}

For numerical stability, all features and stresses are standardized to zero mean and unit variance before training. After inference, the scaling is inverted so that the reported parameters retain physical units. Parameter constraints ($0\le\beta<\alpha\le1$; positive relaxation times) are enforced via smooth sigmoid and softplus reparameterizations, as detailed in Sec.~S3.3. Table~\ref{tab:model_agnostic_decomp} lists four representative memory kernels together with their causal convolutional features; the complete model library, including the constitutive equations and material functions of every model implemented in the repository, is given in Sec.~S4.

\subsection*{S2.1\; Numerical evaluation of the memory features}
\label{sec:SI_feature_eval}

The convolution of Eq.~\eqref{eq:SI_features} extends over $t'\in(-\infty,t]$ and is evaluated in one of two ways, selected according to the deformation history that preceded the record.

\textit{Steady (Fourier-domain) construction.} When the sample can be regarded as having been driven periodically before the record begins, the record of $N$ samples at spacing $\Delta t$ is treated as one period of a periodic protocol, and the features are evaluated exactly at the discrete frequencies of the record, $\omega_k = 2\pi k/(N\Delta t)$ with $k=0,\ldots,N/2$:
\begin{equation}
    \tilde{x}_i(\omega_k) = \tilde{h}_i(\omega_k;\boldsymbol{\kappa})\,\tilde{\gamma}(\omega_k), \qquad \tilde{x}_i(0)=0,
    \label{eq:SI_steady}
\end{equation}
where $\tilde{\gamma}$ is the discrete Fourier transform of the imposed strain and $\tilde{h}_i(\omega;\boldsymbol{\kappa})$ is the analytical transfer function of the corresponding kernel branch, tabulated for every model in Sec.~S4. Because $G^{*}(\omega)=\sum_i c_i\,\tilde{h}_i(\omega)$, the transfer function is the complex modulus of the branch divided by its prefactor. The relaxation kernel itself is never transformed numerically, and the static component is removed. An unpadded $N$-point transform is essential: zero-padding to length $2N$ while using an analytical transfer function implicitly prepends a silent half-period to the periodic history and systematically underestimates the response of long-memory kernels. The construction returns $N$ time-domain samples, so the number of training points seen by the GP is unchanged.

\textit{Causal (time-domain) construction.} When the sample is genuinely at rest before the record begins, the integral is evaluated from $t'=0$ by product integration, in which the singular kernel is integrated analytically over each sampling bin. For a power-law kernel of order $\alpha$,
\begin{equation}
    x(t_n) = \sum_{m=0}^{n} w_m\,\dot{\gamma}(t_{n-m}), \qquad w_m = \frac{\Delta t^{\,1-\alpha}}{\Gamma(2-\alpha)}\left[(m+1)^{1-\alpha}-m^{1-\alpha}\right],
    \label{eq:SI_product_integration}
\end{equation}
with the exponential and Mittag-Leffler kernels, which are regular at $s=0$, sampled at the midpoint of each bin. Integrating the power-law singularity exactly removes the systematic amplitude bias of naive rectangle rules, which reaches ${\sim}15\%$ for $\alpha\simeq0.6$ on typical chirps.

The two constructions differ only in the history they assume before the record, so the choice matters in proportion to how strongly the candidate kernel weights deformation that occurred before the first sample. It is therefore consequential when the memory time is comparable to the record length, and negligible when the record spans many memory times or when the kernel is nearly elastic. We assume that all records analyzed in this work begin from rest, either after a quiescent interval at zero imposed strain or at the start of the acquisition, so the causal construction is used throughout. Nonetheless, the steady construction is offered to the user when the user knows beforehand that the record does not start at rest.

\begin{table*}[tbp]
\caption{Illustrative instantiations of the template $G(t;\Theta)=\sum_{i=1}^{n} c_i\,\phi_i(t;\boldsymbol{\kappa})$ and the associated causal convolutional features $x_i(t)=\int_{0}^{t} \phi_i(t-t';\boldsymbol{\kappa})\,\dot{\gamma}(t')\,\mathrm{d}t'$ that enter the Gaussian processes (GPs). The column $\mathbf{\kappa}$ lists the components that build the kernel's shape and that are optimized jointly with the evidence lower bound. Prefactors $\mathbf{c}$ are recovered as feature-wise GP sensitivities $c_i(t)=\partial\bar{\sigma}(t)/\partial x_i(t)$. Here $E_{a,b}(z)=\sum_{k\ge0} z^{k}/\Gamma(ak+b)$ is the two-parameter Mittag--Leffler function, whose orders $a$ and $b$ and argument $z$ are given in the corresponding kernel entry. The complete model library is given in Sec.~S4.}
\centering
\renewcommand{\arraystretch}{1.6}
\setlength{\tabcolsep}{4pt}
\begin{tabular}{c c c c c c}
\hline
Model &
Model Scheme &
Relaxation Kernel &
Convolutional Feature &
$\mathbf{\kappa}$ &
$\mathbf{c}$ \\
\hline
Maxwell
&
\parbox[c]{1.8cm}{\centering
\includegraphics[width=0.09\textwidth]{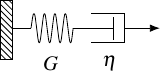}
}
&
\parbox[c]{4cm}{\centering
\rule{0pt}{3.2ex}
$\displaystyle
\begin{aligned}
G(t-t') &= G_c \exp\!\left[-\frac{t-t'}{\tau_c}\right] \\
G_c &= G \\
\tau_c &= \eta / G
\end{aligned}
$
\rule[-1.6ex]{0pt}{0pt}
}
&
\parbox[c]{4.2cm}{\centering
\rule{0pt}{3.2ex}
$\displaystyle
x_1(t)=
\int_{0}^{t}
\exp\!\left[-\frac{t-t'}{\tau_c}\right]
\dot{\gamma}(t')
\,\mathrm{d}t'
$
\rule[-1.6ex]{0pt}{0pt}
}
&
\parbox[c]{0.9cm}{\centering $\tau_c$}
&
\parbox[c]{2.6cm}{\centering
$G_c=
\dfrac{\mathrm{d}\bar{\sigma}(t)}
{\mathrm{d}x_1(t)}$
}
\\
\hline
Scott~Blair
&
\parbox[c]{1.8cm}{\centering
\includegraphics[width=0.09\textwidth]{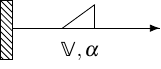}
}
&
\parbox[c]{4cm}{\centering
\rule{0pt}{3.2ex}
$\displaystyle
G(t-t') =
\frac{\mathbb{V}}{\Gamma(1-\alpha)}
(t-t')^{-\alpha}
$
\rule[-1.6ex]{0pt}{0pt}
}
&
\parbox[c]{4.2cm}{\centering
\rule{0pt}{3.2ex}
$\displaystyle
x_1(t)=
\int_{0}^{t}
\frac{1}{\Gamma(1-\alpha)}
(t-t')^{-\alpha}
\dot{\gamma}(t')
\,\mathrm{d}t'
$
\rule[-1.6ex]{0pt}{0pt}
}
&
\parbox[c]{0.9cm}{\centering $\alpha$}
&
\parbox[c]{2.6cm}{\centering
$\mathbb{V}=
\dfrac{\mathrm{d}\bar{\sigma}(t)}
{\mathrm{d}x_{1}(t)}$
}
\\
\hline
\shortstack{Fractional \\ Maxwell}
&
\parbox[c]{1.8cm}{\centering
\includegraphics[width=0.085\textwidth]{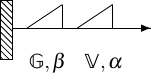}
}
&
\parbox[c]{4cm}{\centering
\rule{0pt}{3.4ex}
$\displaystyle
\begin{aligned}
G(t-t') &= G_c \left(\frac{t-t'}{\tau_c}\right)^{-\beta}
  E_{a,b}(z) \\
a &= \alpha-\beta, \qquad b = 1-\beta \\
z &= -\left(\frac{t-t'}{\tau_c}\right)^{\alpha-\beta} \\
G_c &= \mathbb{V}\tau_c^{-\alpha}\\
\tau_c &= \big(\mathbb{V}/\mathbb{G}\big)^{1/(\alpha-\beta)}
\end{aligned}
$
\rule[-2ex]{0pt}{0pt}
}
&
\parbox[c]{4.2cm}{\centering
\rule{0pt}{3.4ex}
$\displaystyle
x_{1}(t)= \int_{0}^{t}
  \left(\frac{t-t'}{\tau_c}\right)^{-\beta}
  E_{a,b}(z)\,\dot{\gamma}(t')\,\mathrm{d}t'
$
\rule[-2ex]{0pt}{0pt}
}
&
\parbox[c]{0.9cm}{\centering
$\alpha$\\ $\beta$\\ $\tau_c$
}
&
\parbox[c]{2.6cm}{\centering
$G_c=
\dfrac{\mathrm{d}\bar{\sigma}(t)}
{\mathrm{d}x_{1}(t)}$
}
\\
\hline
\shortstack{Fractional \\ Kelvin--Voigt}
&
\parbox[c]{1.8cm}{\centering
\includegraphics[width=0.1\textwidth]{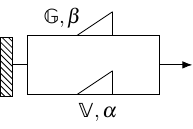}
}
&
\parbox[c]{4cm}{\centering
\rule{0pt}{3.2ex}
$\displaystyle
G(t-t') =
\frac{\mathbb{V}}{\Gamma(1-\alpha)}
(t-t')^{-\alpha}
+
\frac{\mathbb{G}}{\Gamma(1-\beta)}
(t-t')^{-\beta}
$
\rule[-1.6ex]{0pt}{0pt}
}
&
\parbox[c]{4.2cm}{\centering
\rule{0pt}{3.2ex}
$\displaystyle
x_1(t)=
\int_{0}^{t}
\frac{1}{\Gamma(1-\alpha)}
(t-t')^{-\alpha}
\dot{\gamma}(t')
\,\mathrm{d}t'
$\\[2ex]
$\displaystyle
x_2(t)=
\int_{0}^{t}
\frac{1}{\Gamma(1-\beta)}
(t-t')^{-\beta}
\dot{\gamma}(t')
\,\mathrm{d}t'
$
\rule[-1.6ex]{0pt}{0pt}
}
&
\parbox[c]{0.9cm}{\centering
$\alpha$\\ $\beta$
}
&
\parbox[c]{2.6cm}{\centering
$\mathbb{V}=
\dfrac{\partial\bar{\sigma}(t)}
{\partial x_{1}(t)}$ \\[1ex]
$\mathbb{G}=
\dfrac{\partial\bar{\sigma}(t)}
{\partial x_{2}(t)}$
}
\\[10ex]
\hline
\\
\end{tabular}
\label{tab:model_agnostic_decomp}
\end{table*}

\section*{S3.\; Gaussian-process inference}
\label{sec:SI_GP}

\subsection*{S3.1\; Observation model and prior}

For a dataset sampled at times $\{t_j\}_{j=1}^{N}$ with measured stresses $\boldsymbol{\sigma}=[\sigma(t_1),\ldots,\sigma(t_N)]^\top$, we model the stress as
\begin{equation}
    \sigma(t) = f\!\bigl(\mathbf{x}(t;\boldsymbol{\kappa})\bigr) + \varepsilon(t), \qquad \varepsilon(t)\sim\mathcal{N}(0,\mathrm{var}^2),
    \label{eq:SI_gp_obs}
\end{equation}
where $f$ is a latent stress--feature map and $\mathrm{var}^2$ is the observation-noise variance. The noise term $\varepsilon(t)$ captures both instrument noise and model error; it is assumed independent and identically distributed across time points, an assumption examined in Sec.~S5. The latent function is assigned a Gaussian-process prior,
\begin{equation}
    f(\mathbf{x}) \sim \mathcal{GP}\!\bigl[m(\mathbf{x}),\,K(\mathbf{x},\mathbf{x}')\bigr],
    \label{eq:SI_gp_prior}
\end{equation}
where $m(\mathbf{x})=m_0$ is a constant prior mean function (numerically close to zero after feature standardization) and $K(\mathbf{x},\mathbf{x}')$ is the GP covariance kernel, which controls the smoothness of the learned stress--feature map. The overbar notation $\bar{\sigma}(t)$ denotes the posterior predictive mean of $\sigma(t)$ given the training data. The full design matrix is $\mathbf{X}(\boldsymbol{\kappa})\in \mathbb{R}^{N\times n}$, whose $j$th row is $\mathbf{x}(t_j;\boldsymbol{\kappa})^\top$.

\subsection*{S3.2\; Covariance kernels and sensitivity-based prefactor recovery}

We use two GP covariance kernels depending on the analysis goal.

\textit{Linear kernel.}
\begin{equation}
    K_{\mathrm{lin}}(\mathbf{x},\mathbf{x}') = s_f^2\,v_0\,\mathbf{x}^{\top}\mathbf{x}',
    \label{eq:SI_linear_kernel}
\end{equation}
where $s_f^2$ is an output scale and $v_0$ is the variance of the linear base kernel (the implementation composes a scale kernel with a linear base kernel, so the two multiply). With this covariance, the GP is equivalent to Bayesian linear regression in the physics-informed feature space. The posterior predictive mean takes the form $\bar{\sigma}(t)=\sum_i\bar{c}_i\,x_i(t)$, where the overbar denotes a posterior expectation. The feature sensitivities are constant,
\begin{equation}
    \frac{\partial\bar{\sigma}(t)}{\partial x_i(t)} = \bar{c}_i,
    \label{eq:SI_linear_sens}
\end{equation}
and correspond directly to the inferred constitutive prefactors (moduli, viscosities, quasi-properties), each with a posterior credible interval. Using this kernel imposes linearity \textit{a priori}, which is appropriate for protocols designed to operate in the linear viscoelastic regime; it is the kernel used for the wormlike micellar solution, the CNF hydrogel, and the photopolymer resin.

\textit{Radial-basis-function (RBF) kernel.}
\begin{equation}
    K_{\mathrm{RBF}}(\mathbf{x},\mathbf{x}') = s_f^2\exp\!\left[-\frac{\|\mathbf{x}-\mathbf{x}'\|^2}{2\ell^2}\right],
    \label{eq:SI_rbf_kernel}
\end{equation}
where $\ell$ is an isotropic length scale in memory-feature space. This kernel allows a nonlinear stress--feature map. The time-resolved sensitivity,
\begin{equation}
    S_i(t) = \frac{\partial\bar{\sigma}(t)}{\partial x_i(t)},
    \label{eq:SI_rbf_sens}
\end{equation}
is a local slope of the learned map, obtained by automatic differentiation of the posterior mean and sampled over the posterior to give pointwise credible bands. Its approximate constancy over the chirp confirms linear viscoelastic consistency and validates the selected memory kernel; systematic drift in $S_i(t)$ flags model mismatch, nonlinear response, or material mutation. Because this kernel does not presuppose linearity, the constancy of $S_i(t)$ is a test rather than an assumption; it is the kernel used for the UV-curable acrylate.

The scalar prefactor reported for each branch, together with its $95\%$ credible interval, is the mean and the corresponding quantiles taken over the positive-valued posterior samples across the record. Restricting to positive values excludes draws in which the local slope changes sign, which are unphysical for a constitutive prefactor and occur only where the feature amplitude is small.

The capacity of the RBF covariance to represent curvature over the observed data is governed by the ratio of the feature range to $\ell$, both in standardized units: a map that spans many length scales can bend appreciably, whereas one that spans a fraction of a length scale is effectively linear. This ratio is reported alongside the acrylate results in the main text, where it is used to establish that the covariance which admitted the greater curvature is the one that returned a constant sensitivity.

\subsection*{S3.3\; Variational training}

For computational efficiency, we use a sparse variational GP (SVGP)~\cite{hensman2013gaussian,pmlr-v5-titsias09a} with $M$ inducing points $\mathbf{Z}\in\mathbb{R}^{M\times n}$ in memory-feature space. The corresponding inducing values are $\mathbf{u}=f(\mathbf{Z})$. A Gaussian variational distribution
\begin{equation}
    q(\mathbf{u}) = \mathcal{N}(\mathbf{m}_u,\mathbf{S}_u)
    \label{eq:SI_variational}
\end{equation}
is optimized by maximizing the evidence lower bound (ELBO),
\begin{equation}
    \hat{\mathcal{L}} = \mathbb{E}_{q(\mathbf{u})}\!\left[\ln p(\boldsymbol{\sigma}\mid\mathbf{u},\mathbf{X},\mathrm{var}^2)\right] - \mathrm{KL}\bigl[q(\mathbf{u})\,\|\,p(\mathbf{u})\bigr],
    \label{eq:SI_elbo}
\end{equation}
where $p(\mathbf{u})$ is the GP prior evaluated at the inducing inputs. During training, the optimized variables are the variational parameters $(\mathbf{m}_u,\mathbf{S}_u)$, the inducing inputs $\mathbf{Z}$, the GP hyperparameters $(m_0,s_f^2,v_0\ \mathrm{or}\ \ell,\mathrm{var}^2)$, and the viscoelastic shape parameters $\boldsymbol{\kappa}$. The dependence on $\boldsymbol{\kappa}$ enters through the feature matrix $\mathbf{X}(\boldsymbol{\kappa})$, which is re-evaluated at every optimization step (Sec.~S2.1); gradients propagate through the analytical kernels and transfer functions, so the constitutive shape parameters are learned jointly with the GP hyperparameters under the same objective.

The shape parameters are optimized in unconstrained coordinates $\boldsymbol{\rho}$ and mapped to their physical ranges by smooth transformations. Exponents bounded only by $(0,1)$ use the logistic sigmoid $\varsigma(\rho)=[1+e^{-\rho}]^{-1}$, and characteristic times use the softplus $\varsigma_{+}(\rho)=\ln(1+e^{\rho})$,
\begin{equation}
    \alpha = \varsigma(\rho_\alpha), \qquad \beta = \varsigma(\rho_\beta), \qquad \tau_c = \varsigma_{+}(\rho_\tau) + \epsilon,
    \label{eq:SI_reparam}
\end{equation}
with $\epsilon=10^{-4}$ a fixed offset that keeps $\tau_c$ strictly positive. Equation~\eqref{eq:SI_reparam} applies to models carrying a single free exponent, including the Fractional Maxwell liquid used for the micellar solution, in which the remaining exponent is held at $\alpha=0.9999$; the value is displaced from unity by a numerical tolerance so that $\Gamma(1-\alpha)$ remains finite in the relaxation kernel. In models carrying two free exponents (the full Fractional Maxwell and Fractional Kelvin--Voigt families) the ordering $0\le\beta<\alpha\le1$ is enforced by a nested reparameterization,
\begin{equation}
    \alpha = \varsigma(\rho_\alpha), \qquad \beta = \alpha\,\varsigma(\rho_{\Delta}),
    \label{eq:SI_reparam_nested}
\end{equation}
so that $\beta$ depends on both unconstrained coordinates and the constraint holds identically at every optimization step. Optimization uses the Adam optimizer with early stopping on the ELBO. The maximized $\hat{\mathcal{L}}$ serves as the evidence surrogate in model comparison.

\subsection*{S3.4\; Model selection with complete-parameter information criteria}

When the constitutive class is not specified in advance, candidate memories are ranked with the Akaike~\cite{akaike1973second} and Bayesian~\cite{leonard2001bayesian} information criteria,
\begin{equation}
    \mathrm{AIC} = 2\,\mathcal{U} + 2k, \qquad \mathrm{BIC} = 2\,\mathcal{U} + k\ln N,
    \label{eq:SI_bic}
\end{equation}
where $\mathcal{U} = -\hat{\mathcal{L}}$ is the total negative evidence lower bound over the $N$ time samples. The variational objective returns the ELBO normalized per data point; $\mathcal{U}$ is therefore $N$ times the converged per-point bound. Because the ELBO is a lower bound on the log evidence~\cite{pmlr-v5-titsias09a,hensman2013gaussian}, the resulting criteria are conservative surrogates for their exact-likelihood counterparts.

The parameter count $k$ includes all optimized model parameters: (i) the memory-shape parameters $\boldsymbol{\kappa}$; (ii) one scalar per constitutive prefactor $c_i$, inferred through the GP posterior sensitivities; and (iii) a common baseline of four GP hyperparameters, regardless of the kernel choice. For the linear kernel these are the constant mean $m_0$, the output scale $s_f^2$, the linear-kernel variance $v_0$, and the noise variance $\mathrm{var}^2$; for the RBF kernel they are $m_0$, $s_f^2$, the length scale $\ell$, and $\mathrm{var}^2$. Because the baseline is identical for both covariances, model comparisons isolate the competing viscoelastic models. Parameters that are constructed but do not enter the features under the active control mode are excluded: the Fractional Kelvin--Voigt family carries an internal timescale $\tau_c$ that is identified only in stress control, where the creep compliance is a single Mittag--Leffler kernel, and which receives no gradient under strain control. Table~\ref{tab:SI_kcount} lists $k$ for every model in the library.

The variational parameters $(\mathbf{m}_u,\mathbf{S}_u)$, comprising $M+M(M+1)/2$ numbers for $M$ inducing points, are excluded: they parametrize the approximate posterior rather than the model, taking the role of the analytically determined posterior of an exact GP, which contributes no parameters. Their number is identical for all candidates trained with the same configuration, so it cannot alter any ranking.

All candidates compared within a given table are trained on the same record with the same inducing-point count, learning rates, and early-stopping settings. Absolute values of the criteria carry no meaning; only differences with respect to the best-scoring model, $\Delta\mathrm{AIC}$ and $\Delta\mathrm{BIC}$, are interpreted, and the lowest value identifies the most parsimonious constitutive description supported by the time-domain data. The validity of treating the $N$ samples as independent, and the robustness of the resulting rankings, are examined in Sec.~S5.

\begin{table}[htbp]
  \centering
  \caption{Complete parameter count $k$ for the model library. Shape parameters and prefactors are those of Sec.~S4; the four GP hyperparameters are common to all models and both covariance kernels.}
  \label{tab:SI_kcount}
  \renewcommand{\arraystretch}{1.25}
  \begin{tabular}{lccccc}
    \hline
    Model & Shape $\boldsymbol{\kappa}$ & $|\boldsymbol{\kappa}|$ & Prefactors $\mathbf{c}$ & $|\mathbf{c}|$ & $k$ \\
    \hline
    Maxwell      & $\tau_c$                   & 1 & $G_c$                    & 1 & 6 \\
    Scott Blair  & $\alpha$                   & 1 & $\mathbb{V}$             & 1 & 6 \\
    FMG          & $\alpha,\tau_c$            & 2 & $G_c$                    & 1 & 7 \\
    FML          & $\beta,\tau_c$             & 2 & $G_c$                    & 1 & 7 \\
    FMM          & $\alpha,\beta,\tau_c$      & 3 & $G_c$                    & 1 & 8 \\
    FKV          & $\alpha,\beta$             & 2 & $\mathbb{V},\mathbb{G}$  & 2 & 8 \\
    FKV-S        & $\alpha$                   & 1 & $\mathbb{V},G$           & 2 & 7 \\
    FKV-D        & $\beta$                    & 1 & $\eta,\mathbb{G}$        & 2 & 7 \\
    \hline
  \end{tabular}
\end{table}

\section*{S4.\; Model library: constitutive equations and material functions}
\label{sec:SI_library}

This section documents every viscoelastic model implemented in the repository. For each model we give the constitutive equation, the definitions of the characteristic time $\tau_c$ and characteristic modulus $G_c$ where these exist, the relaxation modulus $G(t)$ that generates the causal convolutional feature through Eq.~\eqref{eq:SI_features}, the transfer function $\tilde{h}(\omega)$ used in the steady feature construction of Eq.~\eqref{eq:SI_steady}, and the storage and loss moduli $G'(\omega)$ and $G''(\omega)$. The models follow the conventions of the \texttt{pyRheo} package~\cite{miranda-valdez_niinisto_makinen_lejon_koivisto_alava_2025}. Throughout, $\mathbb{V}$ and $\mathbb{G}$ are quasi-properties with units $\mathrm{Pa\,s}^{\alpha}$ and $\mathrm{Pa\,s}^{\beta}$ respectively~\cite{jaishankar_mckinley_2013, bonfanti_kaplan_charras_kabla_2020}, $G$ is a modulus ($\mathrm{Pa}$), $\eta$ is a viscosity ($\mathrm{Pa\,s}$), $\Gamma(\cdot)$ is the Gamma function, $\delta(\cdot)$ is the Dirac delta, and
\begin{equation}
    E_{a,b}(z) = \sum_{n=0}^{\infty}\frac{z^{n}}{\Gamma(an+b)}, \qquad a>0,
    \label{eq:SI_mittag_leffler}
\end{equation}
is the two-parameter Mittag--Leffler function, which reduces to the exponential for $a=b=1$. The Fractional Maxwell family shares a single kernel form in which the orders are set by the two springpot exponents, $a=\alpha-\beta$ and $b=1-\beta$, so that the gel ($\beta=0$) gives $E_{\alpha,1}$ and the liquid ($\alpha=1$) gives $E_{1-\beta,\,1-\beta}$. The complex modulus follows from $G^{*}(\omega)=i\omega\,\mathcal{F}\{G(t)\}$, with $G^{*}=G'+iG''$, and $\tilde{h}(\omega)$ is the contribution of each branch to $G^{*}$ divided by its prefactor.

\subsection*{S4.1\; Maxwell model (\texttt{Maxwell})}

\textit{Constitutive equation.}
\begin{equation}
    \begin{aligned}
    \sigma(t) + \frac{\eta}{G}\frac{\mathrm{d}\sigma(t)}{\mathrm{d}t} &= \eta\,\frac{\mathrm{d}\gamma(t)}{\mathrm{d}t}, \\
    \tau_c &= \frac{\eta}{G}, \qquad G_c = G .
    \end{aligned}
    \label{eq:SI_const_maxwell}
\end{equation}

\textit{Relaxation modulus and memory feature.}
\begin{equation}
    G(t) = G_c\exp\!\left(-\frac{t}{\tau_c}\right), \qquad x_1(t) = \int_{0}^{t} \exp\!\left(-\frac{t-t'}{\tau_c}\right)\dot{\gamma}(t')\,\mathrm{d}t'.
    \label{eq:SI_relax_maxwell}
\end{equation}

\textit{Transfer function and dynamic moduli.}
\begin{equation}
    \tilde{h}(\omega) = \frac{i\omega\tau_c}{1+i\omega\tau_c}, \qquad G'(\omega) = G_c\frac{(\omega\tau_c)^{2}}{1+(\omega\tau_c)^{2}}, \qquad G''(\omega) = G_c\frac{\omega\tau_c}{1+(\omega\tau_c)^{2}} .
    \label{eq:SI_osc_maxwell}
\end{equation}

\subsection*{S4.2\; Scott~Blair model (\texttt{SpringPot})}

\textit{Constitutive equation.}
\begin{equation}
    \sigma(t) = \mathbb{V}\,\frac{\mathrm{d}^{\alpha}\gamma(t)}{\mathrm{d}t^{\alpha}}, \qquad 0\le\alpha\le1 .
    \label{eq:SI_const_sb}
\end{equation}
The limits $\alpha\to0$ and $\alpha\to1$ recover a Hookean spring of modulus $\mathbb{V}$ and a Newtonian dashpot of viscosity $\mathbb{V}$, respectively~\cite{koeller_1984}.

\textit{Relaxation modulus and memory feature.}
\begin{equation}
    G(t) = \frac{\mathbb{V}}{\Gamma(1-\alpha)}\,t^{-\alpha}, \qquad x_1(t) = \int_{0}^{t} \frac{(t-t')^{-\alpha}}{\Gamma(1-\alpha)}\,\dot{\gamma}(t')\,\mathrm{d}t'.
    \label{eq:SI_relax_sb}
\end{equation}

\textit{Transfer function and dynamic moduli.}
\begin{equation}
    \tilde{h}(\omega) = (i\omega)^{\alpha}, \qquad G'(\omega) = \mathbb{V}\omega^{\alpha}\cos\!\left(\frac{\pi}{2}\alpha\right), \qquad G''(\omega) = \mathbb{V}\omega^{\alpha}\sin\!\left(\frac{\pi}{2}\alpha\right).
    \label{eq:SI_osc_sb}
\end{equation}
The phase angle $\delta = \pi\alpha/2$ and the loss tangent $\tan\delta$ are independent of frequency, which is the defining signature of a critical gel.

\subsection*{S4.3\; Fractional Maxwell gel (\texttt{FractionalMaxwellGel})}

A springpot in series with a Hookean spring. In the shared Fractional Maxwell kernel this is the limit $\beta=0$, so the Mittag--Leffler orders are $a=\alpha$ and $b=1$.

\textit{Constitutive equation.}
\begin{equation}
    \begin{aligned}
    \sigma(t) + \frac{\mathbb{V}}{G}\frac{\mathrm{d}^{\alpha}\sigma(t)}{\mathrm{d}t^{\alpha}} &= \mathbb{V}\,\frac{\mathrm{d}^{\alpha}\gamma(t)}{\mathrm{d}t^{\alpha}}, \\
    \tau_c &= \left(\frac{\mathbb{V}}{G}\right)^{1/\alpha}, \qquad G_c = \mathbb{V}\tau_c^{-\alpha} .
    \end{aligned}
    \label{eq:SI_const_fmg}
\end{equation}

\textit{Relaxation modulus and memory feature.}
\begin{equation}
    G(t) = G_c\,E_{\alpha,1}\!\left[-\left(\frac{t}{\tau_c}\right)^{\alpha}\right], \qquad x_1(t) = \int_{0}^{t} E_{\alpha,1}\!\left[-\left(\frac{t-t'}{\tau_c}\right)^{\alpha}\right]\dot{\gamma}(t')\,\mathrm{d}t'.
    \label{eq:SI_relax_fmg}
\end{equation}

\textit{Transfer function and dynamic moduli.}
\begin{equation}
    \tilde{h}(\omega) = \frac{(i\omega\tau_c)^{\alpha}}{1+(i\omega\tau_c)^{\alpha}},
    \label{eq:SI_h_fmg}
\end{equation}
\begin{equation}
    \begin{aligned}
    G'(\omega) &= G_c\,\frac{(\omega\tau_c)^{2\alpha} + (\omega\tau_c)^{\alpha}\cos\!\left(\frac{\pi}{2}\alpha\right)}{1+(\omega\tau_c)^{2\alpha} + 2(\omega\tau_c)^{\alpha}\cos\!\left(\frac{\pi}{2}\alpha\right)},
    \\
    G''(\omega) &= G_c\,\frac{(\omega\tau_c)^{\alpha}\sin\!\left(\frac{\pi}{2}\alpha\right)}{1+(\omega\tau_c)^{2\alpha} + 2(\omega\tau_c)^{\alpha}\cos\!\left(\frac{\pi}{2}\alpha\right)} .
    \end{aligned}
    \label{eq:SI_osc_fmg}
\end{equation}

\subsection*{S4.4\; Fractional Maxwell liquid (\texttt{FractionalMaxwellLiquid})}

A springpot in series with a Newtonian dashpot. This is the model selected for the wormlike micellar solution in the main text. In the shared Fractional Maxwell kernel this is the limit $\alpha=1$, so the Mittag--Leffler orders are $a=b=1-\beta$.

\textit{Constitutive equation.}
\begin{equation}
    \begin{aligned}
    \sigma(t) + \frac{\eta}{\mathbb{G}}\frac{\mathrm{d}^{1-\beta}\sigma(t)}{\mathrm{d}t^{1-\beta}} &= \eta\,\frac{\mathrm{d}\gamma(t)}{\mathrm{d}t}, \\
    \tau_c &= \left(\frac{\eta}{\mathbb{G}}\right)^{1/(1-\beta)}, \qquad G_c = \eta\,\tau_c^{-1} .
    \end{aligned}
    \label{eq:SI_const_fml}
\end{equation}

\textit{Relaxation modulus and memory feature.}
\begin{equation}
    G(t) = G_c\left(\frac{t}{\tau_c}\right)^{-\beta} E_{1-\beta,\,1-\beta}\!\left[-\left(\frac{t}{\tau_c}\right)^{1-\beta}\right],
    \label{eq:SI_relax_fml}
\end{equation}
\begin{equation}
    x_1(t) = \int_{0}^{t} \left(\frac{t-t'}{\tau_c}\right)^{-\beta} E_{1-\beta,\,1-\beta}\!\left[-\left(\frac{t-t'}{\tau_c}\right)^{1-\beta}\right]\dot{\gamma}(t')\,\mathrm{d}t'.
    \label{eq:SI_feat_fml}
\end{equation}
As $\beta\to0$ the Mittag--Leffler function reduces to the exponential and Eq.~\eqref{eq:SI_relax_fml} recovers the Maxwell relaxation modulus of Eq.~\eqref{eq:SI_relax_maxwell}.

\textit{Transfer function and dynamic moduli.}
\begin{equation}
    \tilde{h}(\omega) = \frac{i\omega\tau_c}{1+(i\omega\tau_c)^{1-\beta}},
    \label{eq:SI_h_fml}
\end{equation}
\begin{equation}
    \begin{aligned}
    G'(\omega) &= G_c\,\frac{(\omega\tau_c)^{2-\beta}\cos\!\left(\frac{\pi}{2}\beta\right)}{1+(\omega\tau_c)^{2(1-\beta)} + 2(\omega\tau_c)^{1-\beta}\cos\!\left(\frac{\pi}{2}(1-\beta)\right)},
    \\
    G''(\omega) &= G_c\,\frac{(\omega\tau_c) + (\omega\tau_c)^{2-\beta}\sin\!\left(\frac{\pi}{2}\beta\right)}{1+(\omega\tau_c)^{2(1-\beta)} + 2(\omega\tau_c)^{1-\beta}\cos\!\left(\frac{\pi}{2}(1-\beta)\right)} .
    \end{aligned}
    \label{eq:SI_osc_fml}
\end{equation}

\subsection*{S4.5\; Fractional Maxwell model (\texttt{FractionalMaxwell})}

Two springpots of orders $\alpha>\beta$ in series~\cite{jaishankar_mckinley_2013,song_holten-andersen_mckinley_2023}. This is the general case of the shared kernel, with Mittag--Leffler orders $a=\alpha-\beta$ and $b=1-\beta$.

\textit{Constitutive equation.}
\begin{equation}
    \begin{aligned}
    \sigma(t) + \frac{\mathbb{V}}{\mathbb{G}}\frac{\mathrm{d}^{\alpha-\beta}\sigma(t)}{\mathrm{d}t^{\alpha-\beta}} &= \mathbb{V}\,\frac{\mathrm{d}^{\alpha}\gamma(t)}{\mathrm{d}t^{\alpha}}, \\
    \tau_c &= \left(\frac{\mathbb{V}}{\mathbb{G}}\right)^{1/(\alpha-\beta)}, \qquad G_c = \mathbb{V}\tau_c^{-\alpha} .
    \end{aligned}
    \label{eq:SI_const_fmm}
\end{equation}

\textit{Relaxation modulus and memory feature.}
\begin{equation}
    G(t) = G_c\left(\frac{t}{\tau_c}\right)^{-\beta} E_{\alpha-\beta,\,1-\beta}\!\left[-\left(\frac{t}{\tau_c}\right)^{\alpha-\beta}\right],
    \label{eq:SI_relax_fmm}
\end{equation}
\begin{equation}
    x_1(t) = \int_{0}^{t} \left(\frac{t-t'}{\tau_c}\right)^{-\beta} E_{\alpha-\beta,\,1-\beta}\!\left[-\left(\frac{t-t'}{\tau_c}\right)^{\alpha-\beta}\right]\dot{\gamma}(t')\,\mathrm{d}t'.
    \label{eq:SI_feat_fmm}
\end{equation}
The Fractional Maxwell gel and liquid are the limits $\beta=0$ and $\alpha=1$ of Eq.~\eqref{eq:SI_relax_fmm}, respectively.

\textit{Transfer function and dynamic moduli.}
\begin{equation}
    \tilde{h}(\omega) = \frac{(i\omega\tau_c)^{\alpha}}{1+(i\omega\tau_c)^{\alpha-\beta}},
    \label{eq:SI_h_fmm}
\end{equation}
\begin{equation}
    \begin{aligned}
    G'(\omega) &= G_c\,\frac{(\omega\tau_c)^{\alpha}\cos\!\left(\frac{\pi}{2}\alpha\right) + (\omega\tau_c)^{2\alpha-\beta}\cos\!\left(\frac{\pi}{2}\beta\right)}{1+(\omega\tau_c)^{2(\alpha-\beta)} + 2(\omega\tau_c)^{\alpha-\beta}\cos\!\left(\frac{\pi}{2}(\alpha-\beta)\right)},
    \\
    G''(\omega) &= G_c\,\frac{(\omega\tau_c)^{\alpha}\sin\!\left(\frac{\pi}{2}\alpha\right) + (\omega\tau_c)^{2\alpha-\beta}\sin\!\left(\frac{\pi}{2}\beta\right)}{1+(\omega\tau_c)^{2(\alpha-\beta)} + 2(\omega\tau_c)^{\alpha-\beta}\cos\!\left(\frac{\pi}{2}(\alpha-\beta)\right)} .
    \end{aligned}
    \label{eq:SI_osc_fmm}
\end{equation}

\subsection*{S4.6\; Fractional Kelvin--Voigt model (\texttt{FractionalKelvinVoigt})}

Two springpots of orders $\alpha>\beta$ in parallel. This is the model selected for the photopolymer resin in the main text. Because the branches act in parallel, their stresses add and the model generates \textit{two} causal features.

\textit{Constitutive equation.}
\begin{equation}
    \begin{aligned}
    \sigma(t) &= \mathbb{V}\,\frac{\mathrm{d}^{\alpha}\gamma(t)}{\mathrm{d}t^{\alpha}} + \mathbb{G}\,\frac{\mathrm{d}^{\beta}\gamma(t)}{\mathrm{d}t^{\beta}}, \\
    \tau_c &= \left(\frac{\mathbb{V}}{\mathbb{G}}\right)^{1/(\alpha-\beta)}, \qquad G_c = \mathbb{V}\tau_c^{-\alpha} .
    \end{aligned}
    \label{eq:SI_const_fkv}
\end{equation}

\textit{Relaxation modulus and memory features.}
\begin{equation}
    G(t) = \frac{\mathbb{V}}{\Gamma(1-\alpha)}t^{-\alpha} + \frac{\mathbb{G}}{\Gamma(1-\beta)}t^{-\beta},
    \label{eq:SI_relax_fkv}
\end{equation}
\begin{equation}
    x_1(t) = \int_{0}^{t} \frac{(t-t')^{-\alpha}}{\Gamma(1-\alpha)}\dot{\gamma}(t')\,\mathrm{d}t', \qquad x_2(t) = \int_{0}^{t} \frac{(t-t')^{-\beta}}{\Gamma(1-\beta)}\dot{\gamma}(t')\,\mathrm{d}t'.
    \label{eq:SI_feat_fkv}
\end{equation}

\textit{Transfer functions and dynamic moduli.}
\begin{equation}
    \tilde{h}_1(\omega) = (i\omega)^{\alpha}, \qquad \tilde{h}_2(\omega) = (i\omega)^{\beta},
    \label{eq:SI_h_fkv}
\end{equation}
\begin{equation}
    \begin{aligned}
    G'(\omega) &= \mathbb{V}\omega^{\alpha}\cos\!\left(\frac{\pi}{2}\alpha\right) + \mathbb{G}\omega^{\beta}\cos\!\left(\frac{\pi}{2}\beta\right),
    \\
    G''(\omega) &= \mathbb{V}\omega^{\alpha}\sin\!\left(\frac{\pi}{2}\alpha\right) + \mathbb{G}\omega^{\beta}\sin\!\left(\frac{\pi}{2}\beta\right).
    \end{aligned}
    \label{eq:SI_osc_fkv}
\end{equation}
The loss contribution of the $\beta$ branch scales as $\mathbb{G}\omega^{\beta}\sin(\pi\beta/2)$, so it vanishes identically at $\beta=0$: a Hookean spring in parallel contributes to $G'(\omega)$ but not to $G''(\omega)$. This is the origin of the qualitative distinction between the FKV and FKV-S descriptions discussed in the main text.

\subsection*{S4.7\; Fractional Kelvin--Voigt-S model (\texttt{FractionalKelvinVoigtS})}

The $\beta=0$ limit of Eq.~\eqref{eq:SI_const_fkv}, in which the second springpot degenerates into a Hookean spring of modulus $G$ acting in parallel with a springpot.

\textit{Constitutive equation.}
\begin{equation}
    \begin{aligned}
    \sigma(t) &= \mathbb{V}\,\frac{\mathrm{d}^{\alpha}\gamma(t)}{\mathrm{d}t^{\alpha}} + G\,\gamma(t), \\
    \tau_c &= \left(\frac{\mathbb{V}}{G}\right)^{1/\alpha}, \qquad G_c = \mathbb{V}\tau_c^{-\alpha} .
    \end{aligned}
    \label{eq:SI_const_fkvs}
\end{equation}

\textit{Relaxation modulus and memory features.}
\begin{equation}
    G(t) = \frac{\mathbb{V}}{\Gamma(1-\alpha)}t^{-\alpha} + G, \qquad x_1(t) = \int_{0}^{t} \frac{(t-t')^{-\alpha}}{\Gamma(1-\alpha)}\dot{\gamma}(t')\,\mathrm{d}t', \qquad x_2(t) = \int_{0}^{t}\dot{\gamma}(t')\,\mathrm{d}t' = \gamma(t),
    \label{eq:SI_relax_fkvs}
\end{equation}
the last equality holding for zero initial strain.

\textit{Transfer functions and dynamic moduli.}
\begin{equation}
    \tilde{h}_1(\omega) = (i\omega)^{\alpha}, \qquad \tilde{h}_2(\omega) = 1,
    \label{eq:SI_h_fkvs}
\end{equation}
\begin{equation}
    G'(\omega) = \mathbb{V}\omega^{\alpha}\cos\!\left(\frac{\pi}{2}\alpha\right) + G, \qquad G''(\omega) = \mathbb{V}\omega^{\alpha}\sin\!\left(\frac{\pi}{2}\alpha\right).
    \label{eq:SI_osc_fkvs}
\end{equation}

\subsection*{S4.8\; Fractional Kelvin--Voigt-D model (\texttt{FractionalKelvinVoigtD})}

The $\alpha=1$ limit of Eq.~\eqref{eq:SI_const_fkv}, in which the first springpot degenerates into a Newtonian dashpot of viscosity $\eta$ acting in parallel with a springpot.

\textit{Constitutive equation.}
\begin{equation}
    \begin{aligned}
    \sigma(t) &= \eta\,\frac{\mathrm{d}\gamma(t)}{\mathrm{d}t} + \mathbb{G}\,\frac{\mathrm{d}^{\beta}\gamma(t)}{\mathrm{d}t^{\beta}}, \\
    \tau_c &= \left(\frac{\eta}{\mathbb{G}}\right)^{1/(1-\beta)}, \qquad G_c = \eta\,\tau_c^{-1} .
    \end{aligned}
    \label{eq:SI_const_fkvd}
\end{equation}

\textit{Relaxation modulus and memory features.}
\begin{equation}
    G(t) = \eta\,\delta(t) + \frac{\mathbb{G}}{\Gamma(1-\beta)}t^{-\beta}, \qquad x_1(t) = \dot{\gamma}(t), \qquad x_2(t) = \int_{0}^{t} \frac{(t-t')^{-\beta}}{\Gamma(1-\beta)}\dot{\gamma}(t')\,\mathrm{d}t'.
    \label{eq:SI_relax_fkvd}
\end{equation}

\textit{Transfer functions and dynamic moduli.}
\begin{equation}
    \tilde{h}_1(\omega) = i\omega, \qquad \tilde{h}_2(\omega) = (i\omega)^{\beta},
    \label{eq:SI_h_fkvd}
\end{equation}
\begin{equation}
    G'(\omega) = \mathbb{G}\omega^{\beta}\cos\!\left(\frac{\pi}{2}\beta\right), \qquad G''(\omega) = \eta\,\omega + \mathbb{G}\omega^{\beta}\sin\!\left(\frac{\pi}{2}\beta\right).
    \label{eq:SI_osc_fkvd}
\end{equation}

\subsection*{S4.9\; Kernel-free baseline (\texttt{KernelFree})}

For completeness, the repository also provides a kernel-free baseline that carries no memory-shape parameters and admits no $G^{*}(\omega)$. Its features are the strain and the strain rate themselves,
\begin{equation}
    x_1(t) = \gamma(t), \qquad x_2(t) = \dot{\gamma}(t),
    \label{eq:SI_kernelfree}
\end{equation}
so that the corresponding sensitivities, $\partial\bar{\sigma}/\partial x_1$ and $\partial\bar{\sigma}/\partial x_2$, are an instantaneous modulus and an instantaneous viscosity. This model has no hereditary structure and is included only as a reference against which memory-carrying candidates can be compared; it is not used in the main text.

\subsection*{S4.10\; Frequency-domain reconstruction}
\label{sec:SI_freq}

After a constitutive model has been selected and its parameters inferred in the time domain, storage and loss moduli are reconstructed by evaluating the analytical expressions of Secs.~S4.1--S4.8 with the inferred parameters. The credible bands reported for the moduli propagate the posterior uncertainty of the prefactors $\mathbf{c}$ only. Because every model in the library is linear in its prefactors, this propagation is exact and requires no linearization: for a single-branch model the relative uncertainty of the modulus equals that of the prefactor at every frequency,
\begin{equation}
    \frac{\mathrm{sd}\left[G'(\omega)\right]}{G'(\omega)} = \frac{\mathrm{sd}\left[G''(\omega)\right]}{G''(\omega)} = \frac{\mathrm{sd}\left[c\right]}{c},
    \label{eq:SI_moduli_var}
\end{equation}
and for the two-branch Fractional Kelvin--Voigt family the band is obtained from the envelope of the moduli evaluated at the credible limits of both prefactors. The memory-shape parameters are held at their point estimates in this reconstruction, so the reported bands are conditional on $\hat{\boldsymbol{\kappa}}$ and should be read as the uncertainty of the modulus scale rather than of the full spectral shape.

This reconstruction is \textit{not} a Fourier transform of the measured chirp signal; it is the frequency-domain consequence of the learned time-domain constitutive model. Within the chirp excitation band $[\omega_1,\omega_2]$, the reconstructed moduli can be compared with discrete Fourier transform (DFT) estimates or independent discrete-frequency-sweep data. Outside the excitation band, the spectra are model-based extrapolations and should be interpreted accordingly.

Equivalently, the trained GP can reach the frequency domain by prediction: evaluating the posterior on an extended chirp and applying the same DFT estimator, $G^{*}(\omega_k)=\tilde{\bar{\sigma}}(\omega_k)/\tilde{\gamma}(\omega_k)$, to the predicted stress. Both routes are shown for the micellar solution in the main text (Fig.~2d), where they coincide with the DFT of the measured stress across the excited band.

\section*{S5.\; Residual diagnostics and the independence assumption}
\label{sec:SI_residuals}

The information criteria of Eq.~\eqref{eq:SI_bic} treat the $N$ time samples as independent observations. This section documents the diagnostics used in the main text to test that assumption and to establish the robustness of the resulting model rankings.

For a fitted candidate we compute the residual sequence $\sigma(t_j)-\bar{\sigma}(t_j)$ and its normalized autocorrelation $C_m$ at lag $m$, where $m$ counts samples as in Eq.~\eqref{eq:SI_product_integration}. The integrated autocorrelation time
\begin{equation}
    \tau_{\mathrm{int}} = 1 + 2\sum_{m\ge1} C_m
    \label{eq:SI_tau_int}
\end{equation}
measures the number of consecutive samples over which the residuals remain correlated; $\tau_{\mathrm{int}}=1$ corresponds to white residuals. The sum is truncated by the initial-positive-sequence rule, in which consecutive lag pairs $C_m+C_{m+1}$ are accumulated while their sum remains positive. The truncation is necessary because the sample autocorrelation of an oscillatory residual continues to ring at high lag, where the estimator is dominated by noise.

Two distinct contributions can produce a nonzero $\tau_{\mathrm{int}}$, and they carry opposite implications. Correlated measurement noise would mean that the record contains fewer independent observations than samples, which would undermine the use of $N$ in Eq.~\eqref{eq:SI_bic}. Correlated model error means only that the candidate kernel leaves a systematic component of the stress unexplained, and does not affect the sampling. The two are separated using the quiescent interval that precedes the sweep in several records (Sec.~S1): there the imposed strain is zero, the memory features vanish identically, and the residual is instrument noise alone, with no model dependence. The autocorrelation time measured over that interval is therefore a property of the instrument, and any excess measured over the sweep is attributable to constitutive misfit.

As a robustness check, the criteria are recomputed after deflating each candidate's evidence term by its own residual autocorrelation time, giving an effective sample size $N_{\mathrm{eff}}=N/\tau_{\mathrm{int}}$ and replacing $N$ by $N_{\mathrm{eff}}$ in the BIC penalty. The correction is deliberately conservative: a residual that is correlated because the kernel is mis-specified is penalized as though the measurement itself carried less information, and the effective sample size differs between candidates, so the comparison is no longer made at a common $N$. It is reported as a stress test of the ordering rather than as a preferred criterion.

\section*{S6.\; Implementation details}
\label{sec:SI_impl}

The framework is implemented in Python using GPyTorch~\cite{gardner2018gpytorch} for GP inference and PyTorch for automatic differentiation and gradient-based optimization. All experiments are run on a single NVIDIA H200 MIG GPU with $71~\mathrm{GB}$ VRAM; training times are typically $<\!3~\mathrm{min}$ per model: $2.23~\mathrm{min}$ for the Fractional Maxwell liquid kernel in the micellar case, $1.11~\mathrm{min}$ for the Scott~Blair kernel in the cellulose nanofiber hydrogel case, and $2.79~\mathrm{min}$ for the Scott~Blair kernel in the acrylate system case.
The number of inducing points is set to $M=300$ for all experiments.

The Mittag--Leffler function of Eq.~\eqref{eq:SI_mittag_leffler} is evaluated for $z\le0$ by a hybrid scheme. For $|z|\le10$ the series is truncated at $K=30$ terms,
\begin{equation}
    E_{a,b}(z) \simeq \sum_{k=0}^{K-1}\frac{z^{n}}{\Gamma(an+b)},
    \label{eq:SI_ml_series}
\end{equation}
and for $z<-10$, where the alternating series becomes numerically unstable through cancellation, the leading term of the asymptotic expansion is used instead,
\begin{equation}
    E_{a,b}(z) \simeq -\frac{1}{z\,\Gamma(b-a)} .
    \label{eq:SI_ml_asymp}
\end{equation}
Both branches are implemented in PyTorch and are differentiable with respect to $a$ and $b$, so gradients of the memory features with respect to the fractional exponents propagate through the Mittag--Leffler evaluation during training.

Code and data supporting this study are openly available at \url{https://github.com/mirandi1/rheogp}.

\begin{acknowledgments}
See main text.
\end{acknowledgments}

\bibliography{supplementary}